\documentclass[10pt]{sigplanconf}

\usepackage{amssymb}
\usepackage{graphicx}

\usepackage{url}

\usepackage{latexsym}
\usepackage{syntonly} 
\usepackage{amsmath}
\usepackage{stmaryrd}
\usepackage{color}
\usepackage{graphics,epsfig,subfigure}

\usepackage{xifthen}
\usepackage{ifpdf}
\usepackage{xkeyval}
\newboolean{usePgfplotsFigures}
\setboolean{usePgfplotsFigures}{true}
\ifthenelse{\boolean{usePgfplotsFigures}}{
  \usepackage{pgfplots}
  \pgfplotsset{compat=1.3}
}{}

\newcommand{\li}{\ar@{-}} 
\newcommand{\lp}{\ar@{.}} 
\newcommand{\fp}{\ar@{.>}} 

\newcounter{todos}
\newcommand{\val}{\mathbb{V}}

\newcommand{\modu}{\ \mbox{\sl mod}\ }

\newcommand{\id}{\mbox{\sl id}}

\newcommand{\Ra}{\Rightarrow}

\def\defi{\mbox{\raisebox{0ex}[1ex][1ex]{$\stackrel{\mbox{\tiny
def}}{\; =\;}$}}}

\newcommand{\comment} [1]{}

\def\Sign{\mbox{\sl Sign\/}}

\def\Par{\mbox{\sl Par\/}}

\def\defemb#1#2{\expandafter\def\csname #1\endcsname
                              {\relax\ifmmode #2\else\hbox{$#2$}\fi}}
 
\def\2c-math#1#2{{\par\medskip\noindent ${#1}$
                      \par\smallskip
                        \noindent\hspace*{\fill} ${#2}$}
                           \\[10pt]}

\def\cal{\mathcal}			   
\defemb{aF}{{\cal F}^\cA}
\defemb{sF}{{\cal F}^{\star}}
\defemb{cA}{{\cal A}}
\defemb{cB}{{\cal B}}
\defemb{cC}{{\cal C}}
\defemb{cD}{{\cal D}}
\defemb{cE}{{\cal E}}
\defemb{cF}{{\cal F}}
\defemb{cG}{{\cal G}}
\defemb{cH}{{\cal H}}
\defemb{cI}{{\cal I}}
\defemb{cJ}{{\cal J}}
\defemb{cK}{{\cal K}}
\defemb{cL}{{\cal L}}
\defemb{cM}{{\cal M}}
\defemb{cN}{{\cal N}}
\defemb{cO}{{\cal O}}
\defemb{cP}{{\cal P}}
\defemb{cQ}{{\cal Q}}
\defemb{cR}{{\cal R}}
\defemb{cS}{{\cal S}}
\defemb{cT}{{\cal T}}
\defemb{cU}{{\cal U}}
\defemb{cV}{{\cal V}}
\defemb{cW}{{\cal W}}
\defemb{cY}{{\cal Y}}
\defemb{cX}{{\cal X}}
\defemb{cZ}{{\cal Z}}
\defemb{cHys}{{\cH_{\cY}^{\cS}}}
\defemb{cHy}{{\cH_{\cY}}}
\defemb{ld}{\dashv}
\defemb{rd}{\vdash}
\defemb{cGH}{{\cal GH}}
\defemb{cGWP}{{\cal GWP}}
\defemb{tL}{{\tt L}}
\defemb{tH}{{\tt H}}
\defemb{tT}{{\tt T}}
\defemb{tS}{{\tt S}}
\defemb{tP}{{\tt P}}
\defemb{tQ}{{\tt Q}}
\defemb{tR}{{\tt R}}
\defemb{tB}{{\tt B\:}}
\defemb{tA}{{\tt A\:}}
\defemb{tC}{{\tt C\:}}
\defemb{tN}{{\tt N}}

\newcommand{\gnsecr}[4]{\mbox{\small $(#1)$}#2\mbox{\small $(#3\Ra #4)$}}

\newcommand{\out}{\mbox{\tt out}}
\newcommand{\ifc}{\mbox{\tt if}}
\newcommand{\thenc}{\mbox{\tt then}}
\newcommand{\elsec}{\mbox{\tt else}}
\newcommand{\nil}{\mbox{\tt skip}}

\newcommand{\while}{\mbox{\tt while}}

\newcommand{\dow}{\mbox{\tt do}}

\defemb{tD}{{\tt D\:}}

\defemb{tX}{{\tt X}}

\newcommand{\f}{\ar@{->}} 

\DeclareSymbolFont{italics}{OT1}{cmr}{m}{it}

\DeclareMathSymbol{a}{\mathalpha}{italics}{"61}
\DeclareMathSymbol{b}{\mathalpha}{italics}{"62}
\DeclareMathSymbol{c}{\mathalpha}{italics}{"63}
\DeclareMathSymbol{d}{\mathalpha}{italics}{"64}
\DeclareMathSymbol{e}{\mathalpha}{italics}{"65}
\DeclareMathSymbol{f}{\mathalpha}{italics}{"66}
\DeclareMathSymbol{g}{\mathalpha}{italics}{"67}
\DeclareMathSymbol{h}{\mathalpha}{italics}{"68}
\DeclareMathSymbol{i}{\mathalpha}{italics}{"69}
\DeclareMathSymbol{j}{\mathalpha}{italics}{"6A}
\DeclareMathSymbol{k}{\mathalpha}{italics}{"6B}
\DeclareMathSymbol{l}{\mathalpha}{italics}{"6C}
\DeclareMathSymbol{m}{\mathalpha}{italics}{"6D}
\DeclareMathSymbol{n}{\mathalpha}{italics}{"6E}
\DeclareMathSymbol{o}{\mathalpha}{italics}{"6F}
\DeclareMathSymbol{p}{\mathalpha}{italics}{"70}
\DeclareMathSymbol{q}{\mathalpha}{italics}{"71}
\DeclareMathSymbol{r}{\mathalpha}{italics}{"72}
\DeclareMathSymbol{s}{\mathalpha}{italics}{"73}
\DeclareMathSymbol{t}{\mathalpha}{italics}{"74}
\DeclareMathSymbol{u}{\mathalpha}{italics}{"75}
\DeclareMathSymbol{v}{\mathalpha}{italics}{"76}
\DeclareMathSymbol{w}{\mathalpha}{italics}{"77}
\DeclareMathSymbol{x}{\mathalpha}{italics}{"78}
\DeclareMathSymbol{y}{\mathalpha}{italics}{"79}
\DeclareMathSymbol{z}{\mathalpha}{italics}{"7A}

\DeclareMathSymbol{A}{\mathalpha}{italics}{"41}
\DeclareMathSymbol{B}{\mathalpha}{italics}{"42}
\DeclareMathSymbol{C}{\mathalpha}{italics}{"43}
\DeclareMathSymbol{D}{\mathalpha}{italics}{"44}
\DeclareMathSymbol{E}{\mathalpha}{italics}{"45}
\DeclareMathSymbol{F}{\mathalpha}{italics}{"46}
\DeclareMathSymbol{G}{\mathalpha}{italics}{"47}
\DeclareMathSymbol{H}{\mathalpha}{italics}{"48}
\DeclareMathSymbol{I}{\mathalpha}{italics}{"49}
\DeclareMathSymbol{J}{\mathalpha}{italics}{"4A}
\DeclareMathSymbol{K}{\mathalpha}{italics}{"4B}
\DeclareMathSymbol{L}{\mathalpha}{italics}{"4C}
\DeclareMathSymbol{M}{\mathalpha}{italics}{"4D}
\DeclareMathSymbol{N}{\mathalpha}{italics}{"4E}
\DeclareMathSymbol{O}{\mathalpha}{italics}{"4F}
\DeclareMathSymbol{P}{\mathalpha}{italics}{"50}
\DeclareMathSymbol{Q}{\mathalpha}{italics}{"51}
\DeclareMathSymbol{R}{\mathalpha}{italics}{"52}
\DeclareMathSymbol{S}{\mathalpha}{italics}{"53}
\DeclareMathSymbol{T}{\mathalpha}{italics}{"54}
\DeclareMathSymbol{U}{\mathalpha}{italics}{"55}
\DeclareMathSymbol{V}{\mathalpha}{italics}{"56}
\DeclareMathSymbol{W}{\mathalpha}{italics}{"57}
\DeclareMathSymbol{X}{\mathalpha}{italics}{"58}
\DeclareMathSymbol{Y}{\mathalpha}{italics}{"59}
\DeclareMathSymbol{Z}{\mathalpha}{italics}{"5A}

\newcommand{\transrel}[1]{\xrightarrow{#1}}
\newcommand{\Actions}{\mathit{Act}}
\newcommand{\Values}{\mathit{Val}}

\newcommand{\trace}{\tau}
\newcommand{\TraceOf}{\mathit{trace}}
\newcommand{\trunc}{\mathit{trunc}}
\newcommand{\Model}{{\cal M}}
\newcommand{\Id}{x}

\newcommand{\Init}{\mathit{init}}

\newcommand{\Epoch}{\mathit{epoch}}
\newcommand{\Len}{\mathit{len}}
\newcommand{\EpistemicEquiv}{\sim}
\newcommand{\True}{\mathit{tt}}
\newcommand{\False}{\mathit{ff}}

\newcommand{\traceOff}[1]{\ensuremath{[\![#1]\!]}}

\newcommand{\ifpresent}[2]{\ifthenelse{\isempty{#1}}{}{#2}}
\newcommand{\languageTerminalFont}[1]{\ensuremath{\text{\texttt{#1}}}}
\newcommand{\Lskip}[0]{\nil{}}
\newcommand{\Lout}[1]{\out{}\ifpresent{#1}{(#1)}}
\newcommand{\Lassign}[2]{\ifpresent{#1}{#1 }\languageTerminalFont{$:=$}\ifpresent{#2}{ #2}}
\newcommand{\Lif}[3]{\ifc{}\ifpresent{#1}{\, #1 \, \Lthen{#2} \, \Lelse{#3}}}
\newcommand{\Lthen}[1]{\thenc{}\ifpresent{#1}{ \, #1}}
\newcommand{\Lelse}[1]{\elsec{}\ifpresent{#1}{ \, #1}}
\newcommand{\Lwhile}[2]{\while{}\ifpresent{#1}{\, #1 \, \Ldo{#2}}}
\newcommand{\Ldo}[1]{\dow{}\ifpresent{#1}{ \, #1}}
\newcommand{\Lseq}[2]{\ifpresent{#1}{#1 \, }\languageTerminalFont{;}\ifpresent{#2}{ \, #2}}
\newcommand{\program}[2][]{\ensuremath{P\ifpresent{#1}{^{#1}}\ifpresent{#2}{_{#2}}}}
\newcommand{\expression}[2][]{\ensuremath{e\ifpresent{#1}{^{#1}}\ifpresent{#2}{_{#2}}}}
\renewcommand{\id}[2][x]{\ensuremath{#1\ifpresent{#2}{_{#2}}}}
\newcommand{\valDom}[0]{\ensuremath{\Values}}
\renewcommand{\val}[2][v]{%
  \ifthenelse{\equal{#1}{true}}{%
    \ensuremath{tt}%
  }{%
    \ifthenelse{\equal{#1}{false}}{%
      \ensuremath{ff}%
    }{%
      \ensuremath{#1\ifpresent{#2}{_{#2}}}%
    }%
  }%
}
\newcommand{\action}[1]{\ensuremath{\alpha\ifpresent{#1}{_{#1}}}}
\newcommand{\execution}[1]{\ensuremath{\pi{#1}}}
\newcommand{\executionPoint}[2][]{\ensuremath{(\execution{#2}, \ifthenelse{\isempty{#1}}{i#2}{#1})}}

\newcommand{\executionPointTrace}[1]{\ensuremath{\tau{#1}}}
\newcommand{\traceOf}[1]{\ensuremath{\TraceOf{}\ifpresent{#1}{(#1)}}}
\newcommand{\model}[2][]{\ensuremath{\Model{}#1\ifpresent{#2}{(#2)}}}
\newcommand{\kripkeStructureSymbol}[0]{\ensuremath{\mathfrak{K}}}
\newcommand{\kripkeStructure}[2][c]{%
  \ifthenelse{\equal{#1}{c}}{%
    \ensuremath{\kripkeStructureSymbol{}\ifpresent{#2}{_{#2}}}%
  }{%
    \ensuremath{(\kripkeStateSet{#2}, \kripkeTruthFct{#2}, \kripkePossibilityRel{#2})}%
  }%
}
\newcommand{\kripkeStateSet}[1]{\ensuremath{S\ifpresent{#1}{_{#1}}}}
\newcommand{\kripkeState}[1]{\ensuremath{s\ifpresent{#1}{_{#1}}}}
\newcommand{\kripkeTruthFct}[1]{\ensuremath{\mathcal{T}\ifpresent{#1}{_{#1}}}}
\newcommand{\kripkePossibilityRel}[1]{\ensuremath{\mathcal{K}\ifpresent{#1}{_{#1}}}}
\newcommand{\interpretedSystemSymbol}[0]{\ensuremath{\mathcal{I}}}
\newcommand{\interpretedSystem}[2][c]{%
  \ifthenelse{\equal{#1}{c}}{%
    \ensuremath{\interpretedSystemSymbol{}\ifpresent{#2}{_{#2}}}%
  }{%
    \ensuremath{(\interpretedSystemRunSet{#2}, \interpretedSystemTruthFct{#2})}%
  }%
}
\newcommand{\interpretedSystemRunSet}[1]{\ensuremath{\mathcal{R}\ifpresent{#1}{_{#1}}}}
\newcommand{\interpretedSystemRun}[1]{\ensuremath{r\ifpresent{#1}{_{#1}}}}
\newcommand{\interpretedSystemTruthFct}[1]{\ensuremath{\mathcal{T}\ifpresent{#1}{_{#1}}}}
\newcommand{\ETLFullLanguage}[0]{\ensuremath{\mathcal{L}_{KU}}}

\newcommand{\ETFormula}[2][1]{\ensuremath{\ifthenelse{\equal{#1}{1}}{\phi}{\psi}\ifpresent{#2}{_{#2}}}}
\newcommand{\operatorFont}[1]{\ensuremath{#1}}
\newcommand{\initialValueOp}[2]{\ensuremath{\Init\ifpresent{#1}{_{#1}}\ifpresent{#2}{(#2)}}}
\newcommand{\knowledgeOp}[2][]{\ensuremath{\operatorFont{K} \ifthenelse{\isempty{#1}}{#2}{(#2)}}}
\newcommand{\untilOp}[3][]{\ensuremath{\ifthenelse{\isempty{#1}}{#2}{(#2)} \operatorFont{U} \ifthenelse{\isempty{#1}}{#3}{(#3)}}}
\newcommand{\possibilityOp}[2][]{\ensuremath{\operatorFont{L} \ifthenelse{\isempty{#1}}{#2}{(#2)}}}
\newcommand{\futureOp}[2][]{\ensuremath{\operatorFont{F} \ifthenelse{\isempty{#1}}{#2}{(#2)}}}
\newcommand{\globallyOp}[2][]{\ensuremath{\operatorFont{G} \ifthenelse{\isempty{#1}}{#2}{(#2)}}}
\newcommand{\weakUntilOp}[3][]{\ensuremath{\ifthenelse{\isempty{#1}}{#2}{(#2)} \operatorFont{W} \ifthenelse{\isempty{#1}}{#3}{(#3)}}}
\newcommand{\releaseOp}[3][]{\ensuremath{\ifthenelse{\isempty{#1}}{#2}{(#2)} \operatorFont{R} \ifthenelse{\isempty{#1}}{#3}{(#3)}}}
\newcommand{\allOp}[3][]{\ensuremath{\forall #2 \ifpresent{#3}{. \: \ifthenelse{\isempty{#1}}{#3}{(#3)}}}}
\newcommand{\existsOp}[3][]{\ensuremath{\exists #2 \ifpresent{#3}{. \: \ifthenelse{\isempty{#1}}{#3}{(#3)}}}}
\newcommand{\impliesOp}[3][]{\ensuremath{\ifthenelse{\isempty{#1}}{#2}{(#2)} \rightarrow \ifthenelse{\isempty{#1}}{#3}{(#3)}}}
\newcommand{\formulaNamesFont}[1]{\ensuremath{\mathsf{#1}}}
\newcommand{\espFormula}[3][c]{%
  \ifthenelse{\equal{#1}{c}}{%
    \ensuremath{\formulaNamesFont{ESP}}%
  }{%
    \allOp[p]{\vec{v}}{%
      \impliesOp{%
        \initialValueOp{#2}{\vec{v}}%
      }{%
        \allOp{\vec{u}}{%
          \possibilityOp[p]{%
            \initialValueOp{#2}{\vec{v}} \land \initialValueOp{#3}{\vec{u}}%
          }%
        }%
      }%
    }%
  }%
}
\newcommand{\espmFormulaRelay}[6]{%
  \ifthenelse{\equal{#1}{true}}{%
    \allOp{\vec{v}_1}{\allOp{\vec{u}_1}{%
      \impliesOp{%
        \initialValueOp{#4}{\vec{v}_1} \land \initialValueOp{#5}{\vec{u}_1}%
      }{%
        \ifthenelse{\equal{#2}{true}}{\hspace{\stretch{1}} \\ \hspace{\stretch{1}}}{}%
        \allOp{\vec{u}_2}{%
          \impliesOp{%
            ( \bigwedge_{#3 \in #6} #3(\vec{v}_1, \vec{u}_1) = #3(\vec{v}_1, \vec{u}_2) ) %
          }{%
            \ifthenelse{\equal{#2}{true}}{\hspace{\stretch{1}} \\ \hspace{\stretch{1}}}{}%
            \possibilityOp[p]{%
              \initialValueOp{#4}{\vec{v}_1} \land \initialValueOp{#5}{\vec{u}_2}%
            }%
          }%
        }%
      }%
    }}%
  }{%
    \ensuremath{\formulaNamesFont{ESPM}\ifpresent{#6}{(#6)}}%
  }%
}
\makeatletter
  \define@boolkeys{espmFormula}[DnifKey@]{expand,cut}[true]
  \define@cmdkey{espmFormula}[dnifKey@]{phi}[\phi]{}
  \presetkeys{espmFormula}{expand=false,cut=false,phi}{}
  \newcommand{\espmFormula}[4][]{%
    \setkeys{espmFormula}{#1}%
    \ifDnifKey@expand\def\dnifKeyExpand{true}\else\def\dnifKeyExpand{false}\fi%
    \ifDnifKey@expand\def\dnifKeyCut{true}\else\def\dnifKeyCut{false}\fi%
    \begingroup%
      \edef\temp{%
        \endgroup%
        \noexpand\espmFormulaRelay%
        {\unexpanded\expandafter{\dnifKeyExpand}}%
        {\unexpanded\expandafter{\dnifKeyCut}}%
        {\unexpanded\expandafter{\dnifKey@phi}}%
        {\unexpanded{#2}}%
        {\unexpanded{#3}}%
        {\unexpanded{#4}}%
      }%
    \temp 
  }
\makeatother
\newcommand{\powersetOf}[1]{\ensuremath{\mathcal{P}\ifpresent{#1}{(#1)}}} 

\newtheorem{definition}{Definition}[section]
\newtheorem{lemma}{Lemma}[section]
\newtheorem{corollary}{Corollary}[section]
\newtheorem{proposition}{Proposition}[section]
\newtheorem{example}{Example}[section]
\newenvironment{proof}{\noindent {\sc Proof.~}}{\hfill 
$\Box$\newline\smallskip\mbox{}\unskip~~\normalsize} 

\def\Idd{\mbox{\sl Id\/}}

\title{Epistemic Temporal Logic for Information Flow Security}

\authorinfo{Musard Balliu \and Mads Dam \and Gurvan Le Guernic}
     {
  Royal Institute of Technology \\
  Stockholm, Sweden }
  {\{musard,mfd,gurvan\}@kth.se}

\newboolean{withLanguageSyntaxFigure} \setboolean{withLanguageSyntaxFigure}{true}
\newboolean{logicSemanticsOnTwoColumns} \setboolean{logicSemanticsOnTwoColumns}{true}
\newboolean{withRelationToStandardModelsOfKnowledge} \setboolean{withRelationToStandardModelsOfKnowledge}{true}

\begin{document}
\conferenceinfo{PLAS'11}{June 5, 2011, San Jose, California, USA.}
\copyrightyear{2011}
\copyrightdata{978-1-4503-0830-4/11/06}


\maketitle

\begin{abstract}
Temporal epistemic logic is a well-established framework for expressing agents knowledge and how it evolves over time. Within language-based security these are central issues, for instance in the context of declassification.  We propose to bring these two areas together. The paper presents a computational model and an epistemic temporal logic used to reason about knowledge acquired by observing program outputs. This approach is shown to elegantly capture standard notions of noninterference and declassification in the literature as well as information flow properties where sensitive and public data intermingle in delicate ways.
\end{abstract}

\category{D.3.1}{Programming Languages}{Formal Definitions and Theory}[Semantics]
\category{F.3.1}{Logics and Meanings of Programs}{Specifying and Verifying and Reasoning about Programs }[Logics of programs]
\category{K.6.5}{Management of Computing and Information Systems}{Security and Protection}

\terms Languages, Security, Verification

\keywords
Information Flow, Epistemic Logic, Noninterference, Declassification

\section{Introduction}\label{sec:intro}

Information flow analysis and language-based security has been a hot topic for well over ten years now. A large array of specification and validation techniques have been proposed, involving security properties (multi-level security, mandatory access control), semantical modeling techniques (trace conditions, simulations and bisimulations/unwinding conditions), and analysis and enforcement techniques (type systems, dependency analyses of various forms). A critique that may be leveled at much of the past work, our own included, is that it has not always managed to separate concerns very clearly. In particular, constraints in specification techniques, programming language features, and details and limitations in the enforcement/analysis mechanisms have been interdependent in such a way that it has often been unclear exactly what properties are enforced and how the various approaches relate to each other. Also, as pointed out by several authors \cite{DBLP:conf/sp/BanerjeeNR08,DBLP:conf/sp/RochaBHWE10}, the policy specification mechanisms have often been interwoven with the object (the program) on which the policy is to be enforced in a manner that makes it hard to separate policy concerns from enforcement concerns. 

A common feature in much recent work on information flow analysis, cf. \cite{ASgr07,DBLP:conf/sp/BanerjeeNR08,DBLP:conf/sp/RochaBHWE10}, has been the appeal to the concept of {\em knowledge} as a fundamental mechanism to bring out what security/confidentiality property is being enforced (the ``revealed" knowledge) and compare it with the knowledge allowed by the policy. This appeal to knowledge, typically as equivalence relations on initial states (or partial equivalence relations \cite{DBLP:conf/esop/SabelfeldS99}), has been important to produce clear, external reference conditions on which e.g. soundness arguments can be based. Knowledge, as it happens, is at the root of an entire branch of logic, namely the logic of knowledge, or epistemic logic.
 In this paper we aim to show that the epistemic logic account of knowledge is compatible with the knowledge notion which has emerged within language-based security, and can have a valuable role to play for policy specification.


Temporal epistemic logic is a well-established framework \cite{FHMV95} which can be used in distributed systems to reason about knowledge and how it evolves over time. Temporal epistemic logic adds epistemic connectives \knowledgeOp[]{} and \possibilityOp[]{} to familiar temporal connectives such as \globallyOp[]{} (always) and \untilOp[]{}{} (until). Those epistemic connectives relate agents local state to the possible global states that are consistent with the agents local observations.
 The property \knowledgeOp[]{\ETFormula[1]{}} expresses that an agent $A$ observing a program ``knows" \ETFormula[1]{} in the sense that \ETFormula[1]{} holds in all states that are possible given $A$'s past observations.
 Dually, \possibilityOp[]{\ETFormula[1]{}} expresses that {\em some} observationally equivalent state exists for which \ETFormula[1]{} holds. Thus, as an example, the property $\ETFormula[1]{} = \globallyOp[p]{C \rightarrow \allOp[]{v}{\possibilityOp[p]{h\!\!=\!\!v}}}$ expresses that whenever some condition $C$ holds then, as far as the attacker can tell, any value of $h$ is possible (and so the value of $h$ is unknown and not released to the attacker). 

In this study we apply temporal epistemic logic to standard sequential while programs augmented with a public output statement, in order to allow a program to "gradually release" \cite{ASgr07} 
 information concerning its initial state. The program model is turned into a model for temporal epistemic logic in the style of interpreted systems \cite{FHMV95}. This is done by defining an S5 perfect recall epistemic accessibility relation using the simple and intuitive idea that two execution states should be regarded as being epistemically the same if they have been reached by identical traces of publicly observable output, i.e. such that an observer cannot tell the two states apart. In particular, if there exists an execution sequence producing a trace $\trace$ and ending in a state refuting property $\phi$ then the attacker  is forced to hold $\lnot\phi$ for possible. 

Our main objective with this paper is to show that temporal epistemic logic is an interesting and natural device with which to express information flow policies for imperative programs. We show this partly by example, and partly by demonstrating how various state-based security conditions related to noninterference 
\cite{goguenmes82,goguen+84uaic} (absence of ``bad'' information flows) 
and declassification \cite{SSJCS07} (intended release of information) can be characterized using the logic. 

We are not the first to apply epistemic logic in the context of computer security. The concrete link between language-based security and temporal epistemic logic which we point out in this paper appears, however, to be new. BAN logic \cite{DBLP:journals/tocs/BurrowsAN90} and successors use epistemic concepts to model agents changing knowledge and belief in security protocols. BAN logic, however, suffered from a lack of an intuitively acceptable semantics (the problem of logical omniscience), something that has only been remedied recently \cite{DBLP:conf/lics/CohenD07}. Post-BAN work in security protocol verification has to a large extent focused on Dolev-Yao types of direct knowledge extraction. This approach works well for many concrete protocols, but it is not adequate to capture the types of indirect channels of high importance in language-based security. For formal analysis of distributed protocols and multi-agent systems, epistemic logic and various extensions have extensive histories \cite{FHMV95}. Much recent work in the area has focused on model checking \cite{Raimondi2007235,DBLP:conf/cav/GammieM04}. Applications of epistemic concepts have been made in process calculi such as the applied $\pi$-calculus \cite{kremer:2009:el4tapc} and CCS \cite{mardare:2006:deohml} and to model protocols for instance in the area  of electronic voting \cite{Baskar:2007:KMV:1324249.1324261}. A precursor of our approach is Askarov and Sabelfeld's gradual release model \cite{ASgr07} where attackers knowledge is modeled as equivalence relations on initial states. In the paper we look into this relationship in more detail and show how gradual release and a number of other possibilistic state-based security conditions can be characterized using temporal epistemic logic.

 In Section~\ref{sec:computational-model} we set up  the underlying computational model. Section~\ref{logic} introduces the syntax and semantics of linear time temporal epistemic logic on these models, and shows how the model relates to the standard interpreted systems model \cite{FHMV95}. We then turn to various well known security conditions from the literature, including noninterference and different flavors of declassification along the dimensions considered by \cite{SSJCS07} in Sect.~\ref{sec:noninterference} to \ref{sec:declassification-when}. We finally point out some open issues and directions for future work.

\section{Computational Model} \label{sec:computational-model}

 In this section we set up our language's basic computational model. We study a simple while language extended with a synchronous output statement that, over the course of a computation, causes information to be leaked to an observer.
 Besides the output statement ``$\Lout{\expression{}}$'', the features of our while language are commonplace: assignments, conditionals, while loops, a primitive data type of values belonging to a finite set $\Values$.
 \ifthenelse{\boolean{withLanguageSyntaxFigure}}{
   The grammar of the language is given in Fig.\ref{fig:grammar}.
 }{}
 Programs are ranged over by \program{}, identifiers by \id{}, values by \val{}, and expressions by \expression{}.

\ifthenelse{\boolean{withLanguageSyntaxFigure}}{
  \begin{figure}[h!tb]
    \begin{center}
      \begin{align*}
        \program{} ::= \ & \Lskip{} \mid \Lout{\expression{}} \mid \Lassign{x}{\expression{}} \mid \Lseq{\program{1}}{\program{2}} \\
        \mid \ & \Lif{\expression{}}{\program{}}{\program{}} \mid \Lwhile{\expression{}}{\program{}}
      \end{align*}
    \end{center}
    \caption{Programming language grammar}
    \label{fig:grammar}
  \end{figure}
}{}

 A store is a finite map $\sigma: \id{} \mapsto \val{}$, and $\sigma(\expression{})$ is the value of \expression{} in store $\sigma$. An execution state is a pair $(\program{},\sigma)$.
 The execution of a program generates observable actions (or events) belonging to $\Actions$ and ranged over by $\alpha$ ($\Actions = \{\Lout{v} \mid v \in \Values\}$).
 The transition relation $(P,\sigma) \transrel{\alpha} (P',\sigma')$, or $(P,\sigma) \transrel{} (P',\sigma')$, states that by taking one execution step in the execution state $(P,\sigma)$ the execution generates the visible event $\alpha$, if it is present, and the new execution state is $(P',\sigma')$. We write $(P,\sigma)\transrel{(\alpha)}(P',\sigma')$ where $\alpha$ is optional. 
\begin{definition}[Execution] \mbox{}\\
An {\em execution} is a finite or infinite sequence of execution states.
\begin{equation}
  \label{101226.1}
  \pi = (P_0,\sigma_0) \transrel{(\alpha_0)}  \cdots \transrel{(\alpha_{n-1})} (P_n,\sigma_n) \transrel{(\alpha_n)}\cdots
\end{equation}
The execution $\pi$ is \emph{maximal} if $\pi$ is a prefix of the execution $\pi'$ only if $\pi = \pi'$.
\end{definition}

 We write $\Len(\pi)$ for the length (number of transitions) of the execution $\pi$. An \emph{execution point}, or simply \emph{point}, is a pair $(\pi,i)$ where $0 \leq i \leq \Len(\pi)$. An execution point $(\pi,i)$ represents the state of the execution $\pi$ after $i$ steps. We write $\trunc(\pi,i)$ for the prefix of $\pi$ up to, and including, execution state $(P_i,\sigma_i)$, the $i$\textsuperscript{th} execution state of $\pi$. We extend the notations as follows: $\pi(i)=(P_i,\sigma_i)$, $P(\pi,i) = P_i$ and $\sigma(\pi,i) = \sigma_i$.

 In our model, the power of the attacker is modeled by providing a function $\TraceOf$ mapping execution points to traces that represent what the attacker has been able to observe so far. In particular, $\TraceOf(\pi,i)$ can span from the truncation function $\trunc(\pi,i)$ for the strongest attacker able to see all the internal computation, to the function returning the last event generated for a weak memory-less attacker. For the standard noninterference attacker able to observe a set of identifiers $X$ during the execution, $\TraceOf$ is the function returning the sequence of stores $\sigma_j$ ($0 \leq j \leq i$) restricted to the domain $X$ and where identical consecutive stores are collapsed. In the remaining of this paper, we use the function $\TraceOf$ given in Def.~\ref{def:trace}. This definition corresponds to the perfect recall attacker, i.e. only able to observe outputs and having memory of past observations.
 \begin{definition}[Trace] \label{def:trace} \mbox{}\\
   A {\em trace} \executionPointTrace{} is an element of $\Actions^*$. \traceOf{}\executionPoint{} is the sequence of events \action{j} such that \mbox{$0 \leq j < i$} and \action{j} exists. The definition of \traceOf{} is trivially extended to executions, such that $\traceOf{\execution{}} = \traceOf{}(\execution{}, \Len(\execution{}))$
 \end{definition}
   The trace of the execution (\ref{101226.1}) is: $(\alpha_0) (\alpha_1) \cdots (\alpha_n) \cdots$

 A model $\Model$ is a set of maximal executions. Normally we take as a model the set of all maximal executions originating from some designated set of initial states, for instance of the shape $(P_0,\sigma_0)$ where $P_0$ is a fixed initial program. We write \model[]{\program{}} for the set of all maximal executions started at all initial states $(\program{},\sigma_0)$ for all initial value stores $\sigma_0$. An {\em epoch} is a set of points reachable by observing a given trace, i.e. $\Model$ is implicit,
 \begin{multline*}
   \Epoch(\trace,\Model) = \\ \{ (\pi,i) \mid \pi \in \Model, 0 \leq i \leq \Len(\pi), \TraceOf(\pi,i) = \trace \}
 \end{multline*}
The epoch of a trace $\trace$ precisely captures the knowledge obtained by observing $\trace$ (in the present possibilistic setting, and ignoring lower level features induced by compilers and run-time systems). For instance, if all points $(\pi,i)\in\Epoch(\trace,\Model)$ have the property that the store at that point assigns to $\Id$ a value between 3 and 5, say, then this fact is known to the observer once she has observed the trace $\trace$. In other words, epoch induce a relations of "equivalent knowledge". Indeed, epochs induce on points a standard epistemic S5 modal accessibility relation $\EpistemicEquiv$ by the condition:
\begin{align*}
                  & \; \executionPoint{} \EpistemicEquiv \executionPoint{'} \\
  \Leftrightarrow & \; (\pi,i)\in\Epoch(\trace,\Model) \text{ implies } (\pi',i')\in\Epoch(\trace,\Model) \\
  \Leftrightarrow & \; \traceOf{}\executionPoint{} = \traceOf{}\executionPoint{'}
\end{align*}

\section{Linear Time Epistemic Logic} \label{logic}
Reflecting the temporal and epistemic structure of models, we propose to use temporal epistemic logic to express
dynamic information flow properties of programs. Many such logics have been considered in the literature \cite{FHMV95}. Here we propose to work with a standard, very general and expressive logic in the family of temporal epistemic logics, namely the linear time temporal epistemic logic \ensuremath{KL_1} without the \textit{Next} operator, in this paper referred to as \ETLFullLanguage{}.

\begin{definition}[Syntax of \ETLFullLanguage{}] \label{110304.1} \mbox{}\\
  The language \ETLFullLanguage{} of formulas $\phi,\psi$ in linear time temporal epistemic logic is given as follows:
  \begin{displaymath} 
    \ETFormula[1]{} , \ETFormula[2]{} 
      ::= \expression{1} = \expression{2} \; \mid \; \Init_\Id(\expression{})
      \; \mid \; \ETFormula[1]{} \land \ETFormula[2]{} \; \mid \: \lnot \ETFormula[1]{}
      \; \mid \; \knowledgeOp[]{\ETFormula[1]{}} \; \mid \; \untilOp[]{\ETFormula[1]{}}{\ETFormula[2]{}}
  \end{displaymath}
\end{definition}
 Besides boolean identities ($\expression{1} = \expression{2}$), the language contains additional atomic propositions $\Init_{\id{}}(\expression{})$ expressing that the value \id{} in the initial state is identical to the value of \expression{} in the current state. The operator \knowledgeOp[]{}
 is the epistemic knowledge operator. \knowledgeOp[]{\ETFormula[1]{}} holds if \ETFormula[1]{} holds in any state equivalent to the current state. In our setting, two states are considered equivalent if the same sequence of outputs has been generated before reaching them. The operator \untilOp[]{}{} is the standard (strong) until operator. The formula \untilOp[]{\ETFormula[1]{}}{\ETFormula[2]{}} holds if \ETFormula[2]{} holds in a future state and \ETFormula[1]{} holds until reaching that state.

 Various connectives are definable in \ETLFullLanguage{} including standard derived boolean operators such as $\lor$ and \impliesOp[]{}{}, 
the truth constants $\True$ and $\False$, universal \allOp[]{\id{}}{} and existential \existsOp[]{\id{}}{} quantifiers over the finite set of values, 
the epistemic possibility operator \possibilityOp[]{\ETFormula[1]{}} meaning that \ETFormula[1]{} holds for at least one epistemically equivalent state, 
the future operator \futureOp[]{\ETFormula[1]{}} requiring \ETFormula[1]{} to eventually hold in the future, 
the ``always" operator \globallyOp[]{\ETFormula[1]{}} meaning that \ETFormula[1]{} holds in any future state, 
and the weak until \weakUntilOp[]{\ETFormula[1]{}}{\ETFormula[2]{}} which does not require \ETFormula[2]{} to eventually hold.
In the remainder of the paper, we use the above connectives as syntactic sugar with the following definitions.

 \begin{definition}[Syntactic sugar \allOp{}{}, \existsOp{}{}, \possibilityOp{}, \futureOp{}, \globallyOp{} and \weakUntilOp{}{}] \mbox{}\\
 \vspace*{2ex}
 \hspace*{\stretch{1}}
 $\displaystyle \allOp[]{\id{}}{\ETFormula[1]{}} = \bigwedge_{\val{} \in \valDom{}} \ETFormula[1]{}[\val{}/\id{}]$
 \hspace*{\stretch{1}}
 $\displaystyle \existsOp[]{\id{}}{\ETFormula[1]{}} =  \bigvee_{\val{} \in \valDom{}} \ETFormula[1]{}[\val{}/\id{}]$
 \hspace*{\stretch{1}}
 \\ \vspace*{2ex}
 \hspace*{\stretch{1}}
 $\displaystyle \possibilityOp[]{\ETFormula[1]{}} = \lnot \knowledgeOp[p]{\lnot \ETFormula[1]{}}$
 \hspace*{\stretch{1}}
 $\displaystyle \futureOp[]{\ETFormula[1]{}} = \untilOp[]{\True}{\ETFormula[1]{}}$ 
 \hspace*{\stretch{1}}
 $\displaystyle \globallyOp[]{\ETFormula[1]{}} = \lnot ( \futureOp[]{\lnot \ETFormula[1]{}} )$ 
 \hspace*{\stretch{1}}
 \\ \vspace*{2ex}
 \hspace*{\stretch{1}}
 $\displaystyle \weakUntilOp[]{\ETFormula[1]{}}{\ETFormula[2]{}} = (\untilOp[]{\ETFormula[1]{}}{\ETFormula[2]{}}) \lor \globallyOp[]{\ETFormula[1]{}}$ 
 \hspace*{\stretch{1}}
\end{definition}

 Since there is no input statement in the programming language, the only way for secrets to enter a computation is through the initial state. This, and also the lack of past-time temporal connectives which would in a more general setting of reactive programs be a natural device to record past inputs, explains the purpose of the initial state predicate $\Init_{\id{}}(\expression{})$ which plays a critical role in capturing what is known "now" of the initial store. It has to be noted that if \expression{} is independent from the current state then, as the initial value of \id{} does not change over time, the majority of temporal variations of $\Init_{\id{}}(\expression{})$ do not change its semantics as long as the computation has not terminated yet ($\Init_{\id{}}(\expression{}) = \futureOp{\Init_{\id{}}(\expression{})} = \globallyOp{\Init_{\id{}}(\expression{})} = \untilOp{\ETFormula[1]{}}{\Init_{\id{}}(\expression{})}$).

 Noteworthy, also, is that outputs are not reflected in the syntax of the logic by corresponding operators or constants. The reason is that output events are of no intrinsic interest to us; they are relevant only in terms of their effect on observer knowledge, of which states are considered equivalent with regard to operators \knowledgeOp[]{} and \possibilityOp[]{}.

\begin{definition}[Satisfaction] \label{def:satisfiability} \mbox{} \\
 \ifthenelse{\boolean{logicSemanticsOnTwoColumns}}{
   Fig.~\ref{fig:satisfiability} defines the
 }{The}
 satisfaction relation $\model[]{}, \executionPoint{} \models \ETFormula[1]{}$ between points in a model $\Model$ and formulas\ifthenelse{\boolean{logicSemanticsOnTwoColumns}}{.}{ is described below.}
 If the model \model[]{} is clear from the context, we write $\executionPoint{} \models \ETFormula[1]{}$ or $\execution{}, i \models \ETFormula[1]{}$ for $\model[]{}, \executionPoint{} \models \ETFormula[1]{}$.
\ifthenelse{\boolean{logicSemanticsOnTwoColumns}}{}{
      \begin{align*}
      \executionPoint{} \models \; & e_1 = e_2
      & \text{iff } & \; \sigma(\pi,i)(e_1) = \sigma(\pi,i)(e_2) \\
      \executionPoint{} \models \; & \Init_\Id(e)
      & \text{iff } & \; \sigma(\pi,0)(\Id) = \sigma(\pi,i)(e) \\
      \executionPoint{} \models \; & \phi \land \psi
      & \text{iff } & \; \executionPoint{} \models \phi \text{ and } \executionPoint{} \models \psi \\
      \executionPoint{} \models \; & \lnot \phi
      & \text{iff } & \; \executionPoint{} \not \models \phi \\
      \executionPoint{} \models \; & \knowledgeOp[]{\ETFormula[1]{}}
      & \text{iff } & \; \forall \execution{'} \in \model[]{}, \forall \executionPoint{'} \in \execution{'} \text{ such that } \\
      &&&& \makebox[0em][r]{$ \traceOf{}\executionPoint{} = \traceOf{}\executionPoint{'} \text{ then } \executionPoint{'} \models \ETFormula[1]{} \;$} \\
      \executionPoint{} \models \; & \untilOp[]{\ETFormula[1]{}}{\ETFormula[2]{}}
      & \text{iff } & \; \exists j : i \leq j \leq \Len(\execution{}), \; \executionPoint[j]{} \models \ETFormula[2]{} \\
      &&&& \makebox[0em][r]{$ \text{ and } \forall k : i \leq k < j , \; \executionPoint[k]{} \models \ETFormula[1]{} \;$}
      \end{align*}
}
  Satisfaction relative to model \model[]{} or program \program{} is:
  \begin{align*}
    \model[]{} \models \; & \ETFormula[1]{} & \text{iff } & \; \forall \execution{} \in \model[]{}, \; \model[]{}, (\execution{},0) \models \ETFormula[1]{} \\
    \program{} \models \; & \ETFormula[1]{} & \text{iff } & \; \model[]{\program{}} \models \ETFormula[1]{}
  \end{align*}
\end{definition}

\ifthenelse{\boolean{logicSemanticsOnTwoColumns}}{
  \begin{figure*}[h!tb]
    \begin{center}
      \begin{align*}
      \model[]{}, \executionPoint{} \models \; & e_1 = e_2
      & \text{iff } & \; \sigma\executionPoint{}(e_1) = \sigma\executionPoint{}(e_2) \\
      \model[]{}, \executionPoint{} \models \; & \Init_\Id(e)
      & \text{iff } & \; \sigma\executionPoint[0]{}(\Id) = \sigma\executionPoint{}(e) \\
      \model[]{}, \executionPoint{} \models \; & \phi \land \psi
      & \text{iff } & \; \executionPoint{} \models \phi \text{ and } \executionPoint{} \models \psi \\
      \model[]{}, \executionPoint{} \models \; & \lnot \phi
      & \text{iff } & \; \executionPoint{} \not \models \phi \\
      \model[]{}, \executionPoint{} \models \; & \knowledgeOp[]{\ETFormula[1]{}}
      & \text{iff } & \; \forall \execution{'} \in \model[]{}, \forall \executionPoint{'} \in \execution{'} \text{ such that } \traceOf{}\executionPoint{} = \traceOf{}\executionPoint{'} , \: \executionPoint{'} \models \ETFormula[1]{} \\
      \model[]{}, \executionPoint{} \models \; & \untilOp[]{\ETFormula[1]{}}{\ETFormula[2]{}}
      & \text{iff } & \; \exists j : i \leq j \leq \Len(\execution{}) \text{ such that }  \executionPoint[j]{} \models \ETFormula[2]{} \text{ and } \forall k : i \leq k < j , \; \executionPoint[k]{} \models \ETFormula[1]{}
      \end{align*}
    \end{center}
    \caption{Formulas satisfaction at execution point}
    \label{fig:satisfiability}
  \end{figure*}
}{}

In terms of epochs the formula \knowledgeOp[]{\ETFormula[1]{}} expresses that \ETFormula[1]{} holds for all points in the current epoch; and, dually, \possibilityOp[]{\ETFormula[1]{}} expresses that \ETFormula[1]{} holds for at least one point in the current epoch, or in other words, that the observer is unable to rule out $\lnot \ETFormula[1]{}$ on the basis of the outputs received so far.

\begin{example}[Basic example]
 If the point \executionPoint{} satisfies the formula \globallyOp[p]{\id{} = 5} then, in all future  execution points of \execution{}, variable \id{} has value 5.
 If  \executionPoint{} satisfies the formula \futureOp[p]{\knowledgeOp[]{\ETFormula[1]{}}} then there exists a point \executionPoint[j]{} (with $j \ge i$) for which \ETFormula[1]{} holds for all points \executionPoint[j']{'} (including \executionPoint[j]{}) having the same trace as \executionPoint[j]{} ($\TraceOf\executionPoint[j]{} = \TraceOf\executionPoint[j']{'}$, i.e. execution \execution{'} after $j'$ steps has generated the same output sequence as execution \execution{} after $j$ steps).
 Combining both previous formulas, if \executionPoint{} satisfies the formula \futureOp[]{\knowledgeOp[]{\globallyOp[p]{\id{} = 5}}} then there exists a trace \executionPointTrace{} of a future point  \executionPoint{} for which \id{} equals 5 in every future point of any point having trace \executionPointTrace{}.
\end{example}

\begin{example}[It is always possible to lose]
 At the program level, if \globallyOp[]{\possibilityOp[]{\futureOp[p]{\id[lost]{} = \val[true]{}}}} for program \program{} then, for all potential traces \executionPointTrace{} of \program{}, there exists an execution of \program{} which at one point has generated the trace \executionPointTrace{} and for which \id[lost]{} will be equal to \val[true]{} at some point in the future. In other words, if the initial state of an execution of \program{} is unknown, whatever output sequence is observed, it is impossible to rule out the fact that losing in the future is still possible.
\end{example}

\begin{example}[Eventually, the initial value is deducible]
 Still at the program level, if \existsOp[]{\val{}}{\futureOp[]{\knowledgeOp[]{\initialValueOp{\id{}}{\val{}}}}} holds for program \program{} then for all executions \execution{} of \program{} there exists a value \val{} and a point \executionPoint{} which generates a trace \executionPointTrace{} for which, for any execution \execution{'} of \program{}, all points \executionPoint{'} generating the same trace \executionPointTrace{} (including \executionPoint{}) are such that the initial value of \id{} is \val{}. In other words, any execution of \program{} will, at some point, have generated an output sequence from which it is possible to deduce the initial value of \id{}.
\end{example}

\ifthenelse{\boolean{withRelationToStandardModelsOfKnowledge}}{

  \subsection{Relation to Standard Models of Knowledge}
 
 Kripke structures are commonly used to give semantics to modal logics, and hence by extension to epistemic logics
 as well  \cite{FHMV95}. A Kripke structure (for a single agent) is a triple \kripkeStructure[e]{} where \kripkeStateSet{} is a set of states, \kripkeTruthFct{} is a valuation assigning to each atomic proposition a predicate on $\kripkeStateSet{}$, and \kripkePossibilityRel{} is a binary accessibility relation on states such that $(\kripkeState{1}, \kripkeState{2}) \in \kripkePossibilityRel{}$ if from the observations made by the observer while in state \kripkeState{1}, it is equally possible to be in state \kripkeState{2}.
    For a given model \model{}, let \kripkeStateSet{\model{}} be the set of all the execution points \executionPoint{} of the executions \execution{} of \model[]{}; let \kripkeTruthFct{\model[]{}} be  the function taking each atomic proposition of the shape  ``$e_1 = e_2$'' or ``$\Init_\Id(e)$''  to the set of points for which the proposition holds according to Def.~\ref{def:satisfiability}; and finally, let \kripkePossibilityRel{\model[]{}} be the binary relation $\EpistemicEquiv$ defined at the end of Sect.~\ref{sec:computational-model}.
   Then \kripkeStructure[e]{\model[]{}} is a Kripke structure for which the standard definitions of the knowledge operators have the same semantics as the one provided in Def.~\ref{def:satisfiability}.
   
    Interpreted systems are a refinement of Kripke structures used to define the semantics of epistemic logics \cite{FHMV95,Raimondi2007235} in terms of multi-agent systems. Roughly, an interpreted system is a pair \interpretedSystem[e]{}, where \interpretedSystemRunSet{} is a set of runs \interpretedSystemRun{} as functions from time to global states. A global state is a tuple composed of an environment state and one state for every agent in the system. Similarly as in the case of Kripke structures, \interpretedSystemTruthFct{} is a function stating if a state formula holds on a given global state.
    For a given model \model{}, let \interpretedSystemRunSet{\model[]{}} be the set of runs \interpretedSystemRun{\execution{}} such that $\execution{} \in \model[]{}$ and $\interpretedSystemRun{}(i)$ is the pair composed of the environment state $\trunc(\execution{},i)$ with actions removed and the agent/attacker state $\TraceOf\executionPoint{}$. Let $\interpretedSystemTruthFct{}$ be defined for formulas of the shape ``$e_1 = e_2$'' or ``$\Init_\Id(e)$'' according to Def.~\ref{def:satisfiability}, as a predicate on global states.
   The semantics of the knowledge operators provided in Def.~\ref{def:satisfiability} is equivalent to their standard semantics over the interpreted system \interpretedSystem[e]{\model[]{}}.

}{}

\section{Noninterference} \label{sec:noninterference}

%

%
We now discuss how the logic applies to information flow security properties, adapted to the present setting
of output-only imperative programs. We first consider the concept of noninterference \cite{goguenmes82}. In a language-based setting and considering a two-level security lattice only, noninterference in a relational (initial-final state) setting requires that no information about initial values of high identifiers (which we want to protect) can flow to final values of low identifiers (which the attacker can observe). This condition is easily adapted to the present setting of output-only programs by instead prohibiting high flow to the public outputs. 

Write $\sigma_1 \approx_{\vec{x}} \sigma_2$ if the two stores $\sigma_1$ and $\sigma_2$ are equivalent with regard to a set of identifiers ${\vec{x}}$, i.e. $\forall x \in \vec{x}. \: \sigma_1(x) = \sigma_2(x)$. Fix now a set of low identifiers $\vec{l}$, and let $\vec{h}$ be its complement, the high identifiers.

\begin{definition}[ONI] \label{def:ONI} \mbox{} \\
 A program \program{} satisfies \emph{output-only noninterference} iff:
 \begin{multline*}
   \forall \pi_1, \pi_2 \in \Model(\program{}). \: \\ \sigma(\pi_1, 0) \approx_{\vec{l}} \sigma(\pi_2, 0) \Rightarrow \TraceOf(\pi_1) = \TraceOf(\pi_2)
 \end{multline*}
\end{definition}
Intuitively, the definition states that there is no information flowing from ${\vec{h}}$ to the attacker if for any maximal execution having trace $\trace$, all maximal executions started with the same values for ${\vec{l}}$ produce the same trace. In other words, all initial secret values (${\vec{h}}$) might have given rise to the output sequence that an attacker is observing. It is worth noting that this definition subsumes standard noninterference. Indeed, we only need to modify program $P$ by outputting the values of low identifiers (${\vec{l}}$) whenever they are observable. Termination sensitivity can also be added by a final dummy output. We now show how ONI can be encoded in our epistemic framework.
\begin{definition}[\espFormula{}{}] \label{ESP-formula} 
  \begin{displaymath}
    \espFormula{\vec{l}}{\vec{h}} \; \defi \; \espFormula[e]{\vec{l}}{\vec{h}}
  \end{displaymath}
\end{definition}
The formula \espFormula{\vec{l}}{\vec{h}} is satisfied at a given execution point if every initial secret is possible among the execution points having the same trace and initial public values. In our epistemic framework, we claim that a program does not reveal any secret if all its execution points satisfy \espFormula{\vec{l}}{\vec{h}}, i.e. every initial secret is possible for every trace and public inputs generating such trace.
\begin{definition}[AK] \label{def:AK} \mbox{} \\
 A program \program{} satisfies \emph{absence of knowledge} iff:
  \begin{displaymath}
    \program{} \models \; \globallyOp[p]{\espFormula{\vec{l}}{\vec{h}}}
  \end{displaymath}
\end{definition}   
We first give some examples to show how the logic applies to programs wrt. standard noninterference and afterwards prove the equivalence of the above definitions.
\begin{example}
  Let $P::= x:=y;\Lout{y}$ be a program over booleans with $x \in {\vec{h}}, y\in {\vec{l}}$. Then $P$ satisfies ONI since the initial value of $y$ never changes.  We show that $P$ satisfies AK.
  Consider a model $\Model$ associated with program $P$ where the store is a pair $(x,y)$. Then
  \[\Model::=  \left \{ \begin{array}{l}   
      \pi_1=(\True, \True) \stackrel{}{\rightarrow} (\True, \True) \stackrel{\True}{\rightarrow} (\True, \True) \\
      \pi_3=(\True, \False) \stackrel{}{\rightarrow} (\False, \False) \stackrel{\False}{\rightarrow} (\False, \False) \\
      \pi_2=(\False, \False) \stackrel{}{\rightarrow} (\False, \False) \stackrel{\False}{\rightarrow} (\False, \False)\\
      \pi_4=(\False, \True) \stackrel{}{\rightarrow} (\True, \True) \stackrel{\True}{\rightarrow} (\True, \True)
    \end{array} \right. \]
  One can verify, by  case analysis, that  $\Model \models \globallyOp[p]{\espFormula[c]{y}{x}}$. Consider  for instance $\pi_4$. Then $v=\True$ and $\pi_4, i \models \Init_y(v)$ holds for all $0 \leq i \leq 2$. 
  We show that $\pi_4, i \models \forall  u.  L (\Init_{y}(v) \wedge  \Init_{x}(u))$ for all $i$. For $i\in \{0,1\}$, $\TraceOf(\pi_4, i)= \epsilon$, so we can find $(\pi_1,0)$ and $(\pi_2,0)$ if $u=\True$ and $u=\False$, respectively. If $i=2$ and $u = \True$, then $(\pi_1, 2)$ 
  has the same trace and initial value; otherwise, if $u= \False$, we pick $(\pi_4, 2)$. Similarly, the condition holds for other cases. \\
  Let now $P::= x:=y;\Lout{y}$ with $x\in {\vec{l}}, y \in {\vec{h}}$. Then, $P$ falsifies ONI since we output the secret value $y$ to public output. We show for model $\Model$  that
  $\Model \not\models \globallyOp[]{} \; \forall v. (\Init_{x}(v) \rightarrow \forall  u .  L (\Init_{x}(v) \wedge  \Init_{y}(u)))$ i.e. $\exists \pi. \exists i. \exists v.  \Init_x(v) \wedge \exists u. \forall \pi'. \\ \forall i'. \TraceOf(\pi,i) = \TraceOf(\pi',i')$ then 
  $\pi', i' \not \models (\Init_{x}(v) \wedge  \Init_{y}(u)))$. In particular,  $\pi_3$ is a counterexample. Set $v=\True$ and $u=\True$; the only executions having the same trace as $\pi_3$ are $\pi_2$ and $\pi_3$. However,
  $\sigma(\pi_2, 0)(x) = \False \neq v$ and $\sigma(\pi_3, 0)(y) = \False \neq u$.    
\end{example}

\begin{lemma}[Initial values stability] \label{box} \mbox{} 
 For all vectors of values $\vec{v}$ and identifiers $\vec{x}$:
 \begin{displaymath}
   \execution{},0 \models \Init_{\vec{x}}(\vec{v}) \text{ implies } \forall \executionPoint{} \in \execution{} : \; \execution{},i \models \Init_{\vec{x}}(\vec{v})
 \end{displaymath}
 \begin{proof}
 Immediate. By definition of satisfaction relation $\pi, i \models \Init_{\vec x}(\vec v)$  iff $\sigma(\pi,0)(\vec x) = \vec v$.  
 \end{proof}
\end{lemma}

\begin{proposition}[Equivalence of ONI and AK] \label{prop:eq-oni-ak}
  For all programs \program{}:
  \begin{displaymath}
    \program{} \models \text{ONI} \; \text{ iff } \; \program{} \models \text{AK}
  \end{displaymath}
  \begin{proof}
    $(\Rightarrow)$ Assume $P$ satisfies ONI. By definition, given $\pi_1$, then for all $\pi_2 $. $\sigma(\pi_1, 0) \approx_{\vec{l}} \sigma(\pi_2, 0)$,  $\TraceOf(\pi_1) = \TraceOf(\pi_2)$.
    In particular any two equal traces have equal prefix traces of same length. We show that 
    $\pi \in \Model$. $\pi, 0 \models \; \globallyOp[]{} \; \forall \vec{v}.(\Init_{\vec l}(\vec{v}) \rightarrow \forall \vec{u}.  L (\Init_{\vec l}(\vec{v}) \wedge  \Init_{\vec h}(\vec{u})))$. 
    Pick any $\pi \in \Model$ and $\vec{v} \in Val$; then we show for all $0\leq i \leq \Len(\pi)$. $\pi, i \models  (\Init_{\vec l}(v) \rightarrow \forall \vec{u}.  L (\Init_{\vec l}(v) \wedge  \Init_{\vec h}(\vec{u})))$.
    Namely, assume $\pi, i \models \Init_{\vec l}(\vec{v})$ then for any $\vec{u}\in Val$ there exists $\pi', i'$.  $\TraceOf(\pi,i) = \TraceOf(\pi',i') \wedge  (\pi', 0)\models \Init_{\vec l}(\vec{v}) \wedge \Init_{\vec h}(\vec{u})$. 
    Let now $\Model_a \subseteq \Model$ be such that $\forall \pi\in \Model_a$. $\sigma(\pi, 0)(l) = a$. Then $\Model = \bigcup_{a \in Val} \Model_a$ . By ONI condition, for all $\pi \in \Model_a$. $\TraceOf(\pi) = \trace$ for some trace $\trace$ and 
    any initial $\vec h$. Then, using Lemma \ref{box} and  chopping off execution $\pi$ we get the result for all $(\pi,i$). 
    The same argument can be used for any $\Model_a$, so we are done.\\
    $(\Leftarrow)$ Suppose  now $ \forall \pi \in \Model$. $\pi, 0 \models \; \globallyOp[]{} \;\forall \vec{v}.  (\Init_{\vec l}(v) \rightarrow \forall \vec{u}.  L (\Init_{\vec l}(v) \wedge  \Init_{\vec h}(\vec{u})))$. We show ONI holds. 
    By hypothesis, pick  $\pi \in \Model$ with  $\sigma(\pi,0)(l)=v$, then we show that for all $\pi'$ such that $\sigma(\pi',0)(l) = v$, $\TraceOf(\pi)= \TraceOf(\pi')$. By hypothesis, given $\pi$, in particular it is always possible 
    to find $\pi'$ with same initial values $\vec{v}$, for any $\vec{u}$ having the same trace. 
  \end{proof}
\end{proposition}

\begin{example}
 Let $P$ be a program manipulating two private variables $h_1, h_2$ over boolean domain.
 \[ P::=  \left. \begin{array}{ll}
 \ifc \; h_1 \; \thenc \; \Lout{\neg h_2} \; \elsec \;  \Lout{h_2}&
\end{array} \right. \]
The program is not secure since it reveals whether the secrets are equal or not i.e. $h_1 = h_2$. In fact, for all input states where $h_1 = h_2$ i.e. $(\True, \True), (\False, \False)$, $P$ outputs $\False$, otherwise it outputs $\True$ and this is captured by Def.~\ref{def:AK}.
\end{example}
On the other hand, we will see in the following section that if one agrees to declassifies $\phi:= h_1 = h_2$ then Def.~\ref{def:AKD} will deem the program secure.

\section{Declassification: What}

Noninterference guarantees an end-to-end confidentiality policy, namely  as soon as a program conveys 1 bit of secret information, it is ruled out by the condition. In real  applications this policy turns out to be restrictive, as in many  scenarios partial information leakage is considered admissible. Declassification policies  handle those acceptable, or even desired, information leakages \cite{SSJCS07}.
For example, a customer may be allowed to access a scientific article (secret data)
once she has paid the registration fee to some on line provider. In this case, an intentional release of secret information is needed. Declassification has been 
recognized as one of the main challenges in information flow security \cite{SM03}. The main concern is to prove that declassification is safe and the attacker is unable to compromise the release mechanism and disclose more sensitive
information than stated in the policy. Many authors have addressed the problem  from different points \cite{cohen78,SM04,MS04,ASgr07,BCR08,BM10}. In particular, in \cite{SSJCS07}, the authors present a classification of
different flavors of declassification. In this section and the following ones, we show how our temporal epistemic framework captures in an elegant way those dimensions.

One way of modeling declassification is by means of a predicate $\phi$ over initial values which expresses the property one intends to declassify. In that case, one has to make sure that states having the same property $\phi$ can not be distinguished by the attacker. This idea originates from \textit{selective dependency} \cite{cohen78} and corresponds to the \textit{What} dimension \cite{SSJCS07}.
In particular, the programmer should specify a global declassification policy
$\phi$ and the enforcement mechanism has to ensure that no information other than what is specified  in the policy can be disclosed by the attacker. For example, the information system of a company can release the average salary of an employee, but it shouldn't be possible to 
reveal, for instance, the salary of a certain employee. Let $\sigma_1 \approx_{\phi} \sigma_2$ denote equivalent states according to the declassification policy $\phi$ i.e.  $\sigma_1(\phi) = \sigma_2(\phi)$.
 \begin{definition}[NID] \label{def:NID} \mbox{}\\
   Let $\phi$ be a global declassification policy.
   A program \program{} satisfies \emph{noninterference modulo declassification} $\phi$ iff:
   \begin{multline*}
     \forall \pi_1, \pi_2 \in \Model(\program{}). \: \\
     ( \sigma(\pi_1, 0) \approx_{{\vec{l}}} \sigma(\pi_2, 0)  \land \sigma(\pi_1, 0) \approx_{\phi} \sigma(\pi_2, 0) ) \\
     \Rightarrow \TraceOf(\pi_1) = \TraceOf(\pi_2)
   \end{multline*}
 \end{definition}
 The definition of NID specifies that any initial state having the same public values and agreeing on $\phi$ should produce the same output trace.

 Let us now see how global declassification policies can be expressed in our model.
 We first introduce the formula \espmFormula{}{}{}. An execution point satisfies \espmFormula{\vec{l}}{\vec{h}}{\Phi} where $\Phi$ is a set of declassification policies iff, among the other execution points having the same trace and initial public values, every initial secret agreeing on $\Phi$ is possible.

 \begin{definition}[\espmFormula{}{}{}] \label{def:ESPM-formula} 
  \begin{multline*}
    \espmFormula{\vec{l}}{\vec{h}}{\Phi}\; \defi \; \\
    \hspace{1em} \espmFormula[expand,cut]{\vec{l}}{\vec{h}}{\Phi}
  \end{multline*}
 \end{definition}
 
 \begin{proposition}[Equivalence of \espFormula{\vec{l}}{\vec{h}} and \espmFormula{\vec{l}}{\vec{h}}{\emptyset}] \label{prop:ESPf-eq-ESPMf-empty}
   For all execution points \executionPoint{}:
   \begin{displaymath}
     \executionPoint{} \models \espFormula{\vec{l}}{\vec{h}} \; \text{ iff } \; \executionPoint{} \models \espmFormula{\vec{l}}{\vec{h}}{\emptyset}
   \end{displaymath}
   \begin{proof}
     This proposition follows directly from the fact that if $\Phi$ is empty then $\bigwedge_{\phi \in \Phi}$ is vacuously true and \initialValueOp{\vec{h}}{\vec{u}_1} holds for at least one vector of values $\vec{u}_1$.
   \end{proof}
 \end{proposition}

 \begin{proposition}[Monotonicity of \espmFormula{}{}{}] \label{prop:ESPMf-monotonicity}
   For all execution points \executionPoint{} and sets of declassifications $\Phi$ and $\Psi$:
   \begin{displaymath}
     \executionPoint{} \models \espmFormula{\vec{l}}{\vec{h}}{\Phi} \; \text{ implies } \; \executionPoint{} \models \espmFormula{\vec{l}}{\vec{h}}{\Phi \cup \Psi}
   \end{displaymath}
   \begin{proof}
     This proposition follows trivially from the second implication in the formula of \espmFormula{}{}{}. Whenever the left part of the implication $\bigwedge_{\phi \in \Phi \cup \Psi}$ holds then $\bigwedge_{\phi \in \Phi}$ also holds; and the right part of the implication is the same in both cases, so if the \possibilityOp[]{} formula holds with $\Phi$ it still holds with $\Phi \cup \Psi$.
   \end{proof}
 \end{proposition}

 \begin{corollary}[\espFormula{}{} subsumes \espmFormula{}{}{}] \label{cor:ESPf-subsumes-ESPMf}
   For all execution points \executionPoint{} and sets of declassifications $\Phi$:
   \begin{displaymath}
     \executionPoint{} \models \espFormula{\vec{l}}{\vec{h}} \; \text{ implies } \; \executionPoint{} \models \espmFormula{\vec{l}}{\vec{h}}{\Phi}
   \end{displaymath}
   \begin{proof}
     This is a direct corollary of Prop.~\ref{prop:ESPf-eq-ESPMf-empty} and~\ref{prop:ESPMf-monotonicity}.
   \end{proof}
 \end{corollary}

 \begin{definition}[AKD] \label{def:AKD} \mbox{}\\
   Let $\phi$ be a global declassification policy.
   A program \program{} satisfies \emph{absence of knowledge modulo declassification} $\phi$ iff:
   \begin{displaymath}
     \program{} \models \globallyOp[p]{ \espmFormula{\vec{l}}{\vec{h}}{\{\phi\}} }
   \end{displaymath}
 
 \end{definition}

\begin{figure}[htbp]
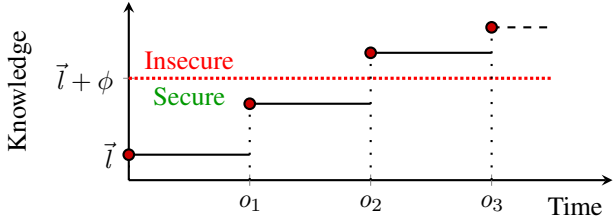

\begin{center}
 
\ifpdf
  \ifthenelse{\boolean{usePgfplotsFigures}}{
    \begin{tikzpicture}
      \begin{axis}[compat=1.3,
        width=.95\linewidth, height=25ex, 
        axis x line = bottom, axis y line = left, 
        xmin=0, enlarge x limits={abs=1,upper},
        ymin=0, enlarge y limits={abs=1,upper},
        xlabel=\text{Time}, every axis x label/.style={at={(ticklabel cs:1)},anchor=south east}, 
        ylabel=\text{Knowledge}, 
        every axis/.append style={thick},
        xtick=\empty, ytick=\empty,
        extra x ticks={2,4,6}, extra x tick labels={$o_1$,$o_2$,$o_3$},
        extra y ticks={1,4}, extra y tick labels={$\vec{l}$,$\vec{l} + \phi$},
      ]
    
      \addplot[red, very thick, densely dotted] coordinates {(0,4) (7,4)};
    
      \addplot+[black, mark=*, jump mark left] coordinates {
        (0,1) (2,3) (4,5) (6,6)
      };
      \addplot+[black, no markers, dashed] coordinates {(6,6) (7,6)};
      \addplot+[black, loosely dotted, no markers, ycomb] coordinates {
        (0,1) (2,3) (4,5) (6,6)
      };
   
      \node[red] at (axis cs:1,4.7) {Insecure};
      \node[green!60!black] at (axis cs:1,3.3) {Secure};
   
      \end{axis}
    \end{tikzpicture}
  }{
    \resizebox{.95\linewidth}{19ex}{\includegraphics{release1.pdf}}
  }
\else
 \scalebox{0.6}{ \input{release1.pstex_t} } 
\fi

\caption{Knowledge and Declassification}
\label{figure:example1}
\end{center}
\end{figure}
Figure~\ref{figure:example1} illustrates the intuition behind our security condition. The graphic presents the knowledge about initial secrets that an attacker gains by observing a certain trace $\trace = o_1o_2o_3$ as function of time elapsed from the beginning of computation.   
The black solid line shows the evolution of attacker knowledge at each output point and in particular how it can possibly increase in each epoch. Initially the attacker has knowledge about public identifiers.
On the other hand the red dotted line shows the global declassification policy represented by a predicate $\phi$. As long as the solid line  remains
below the dotted line the declassification is safe, namely the attacker knowledge is  smaller than the information released intentionally prior to program execution.  In this case, one can see that after the second
observation point $o_2$ the attacker learns more than the policy allows, thus the program becomes insecure.
 \begin{proposition}[Equivalence of NID and AKD] \label{sec:onidak}
  For all programs \program{}:
  \begin{displaymath}
    \program{} \models \text{NID} \; \text{ iff } \; \program{} \models \text{AKD}
  \end{displaymath}
  \begin{proof}
    The proof is similar to the one for Prop.~\ref{prop:eq-oni-ak}.
  \end{proof}
 \end{proposition}
 It is worth noting that if  the declassification policy  states ``\textit{No secret information can be leaked}'', then the property becomes $\phi = \True$  and AKD will correspond to AK. We illustrate the above condition by means of an example.
\begin{example}
 Consider the program $P$  with $h\in{\vec{h}}$. 
\begin{center}
 $P ::= \ifc\; (h=0) \; \thenc\; \Lout{1} \;\elsec\; \Lout{2}$
\end{center}
One can spot an implicit flow due to dependence on a conditional on secret $h$. Let $\Model$ be a model of $P$. To falsify Def.~\ref{def:AK}, pick  $\pi$ such that $\sigma(\pi,0)(l) = \sigma(\pi,0)(h) = 0$. Then, pick $\pi'$ such that $\sigma(\pi',0)(l)=0$ and $\sigma(\pi',0)(h)\neq0$. It is easy to see that $\TraceOf(\pi) \neq \TraceOf(\pi')$. Suppose now we declassify the zeroness of $h$ i.e. $\phi := (h=0)$. All executions originating from $h=0$ produce the same trace i.e. output 1. On the other hand, all executions originating from $\neg \phi := (h \neq 0)$ also produce the same trace, i.e. output 2. Hence, the program is secure.
 It is worth to noting how Def.~\ref{def:AKD} rules out programs that reveal more than what is allowed by the declassification policy. Suppose we want to declassify the sign of identifier $h$, namely $\phi := (h \geq 0)$. Then, $P$ becomes insecure since the attacker is now able to distinguish between values having the same property $\phi$. In particular let $h_1=0$ and $h_2=1$, so $\phi(h_1) = \phi(h_2)$. In that case $P$ outputs $1$ and $2$, respectively, so it is deemed insecure. 
\end{example}
\paragraph*{Abstract Non-Interference}
Abstract Non-Interference (ANI) is an abstract interpretation based approach for modeling and certifying information flow properties\cite{GM04popl}. This  framework characterizes different qualitative aspects related to global declassification policies and attacker observational  power. 
In particular, using the notion of abstract domain, the authors give an extensional model of what an  attacker is allowed to see of public data (attacker power) and of what she is allowed to disclose of secret data (declassification). 
For example, let $P$ be a program with $l\in {\vec{l}}, h \in {\vec{h}}$. 
\[P::= \ifc\; (h\geq0) \; \thenc\; l:=2l\ast h \;\elsec\; l:=2l\ast h+1\]
Clearly, there is an direct flow to public identifier $l$ which conveys the value of secret $h$. However, if one is interested in releasing only 
the sign of secret identifier $h$ in input and  considers a weaker  attacker who is able to observe only the parity of identifier $l$ in output then $P$ will be secure. Indeed, fix the initial value of low identifier $l$  and consider initial values of $h$
in input having the same sign, say $h< 0$. It can be easily seen that the final value of $l$ will have the same parity; in this case it will correspond to an odd value. This definition is called \textit{Narrow ANI via allowing} \cite{M05}.
Let $\eta,\phi, \rho$ be the abstract domains for public input, declassified private input and public output, respectively.

\begin{definition}[NANI] \mbox{}\\
A program \program{} satisfies \emph{Narrow ANI}, ${\gnsecr\eta P \phi \rho}$, iff:
\begin{multline*}
  \forall l_1,l_2 \in {\vec{l}}, \forall h_1, h_2 \in {\vec{h}} : \\
  \eta(l_1) = \eta(l_2) \wedge \phi(h_1) = \phi(h_2) \\
  \Rightarrow \rho(\traceOff{ P}(h_1, l_1)) = \rho(\traceOff{P}(h_2, l_2))
\end{multline*}
\end{definition}

Basically it states that for any initial public values having property $\eta$ and for any private initial values having property $\phi$, the result of the computation  has property $\rho$ over public outputs.
In particular the previous example corresponds to checking ${\gnsecr\Idd P \Sign \Par}$. \\
There is a nice relation between NANI and our epistemic framework. One can look at the abstractions over public input domain and public output domain as abstractions over  channels receiving and releasing  
these values, respectively. More concretely, suppose one wants to check NANI for  ${\gnsecr\eta P \phi \rho}$. In order to model the attacker power in output we can use the output actions $\Lout{e}$ and check the following formula wrt. a model $\Model$
of the program $P;\Lout{\rho(l)}$. Given a pair $(\vec u, \vec v)$ we denote by $\mathop{fst}$ and $\mathop{snd}$, respectively, the first and the second component of such a pair.

\begin{definition}[AAK]\label{sec:aak} \mbox{} \\
  A program \program{} satisfies \emph{abstract absence of knowledge} w.r.t.  abstractions $\rho$, $\eta$ and $\phi$ iff:
  \begin{align*}
    \Lseq{\program{}}{\Lout{\rho(\vec{l})}} \models \globallyOp[p]{ \espmFormula{\vec{l}}{\vec{h}}{\{\eta \circ \mathop{fst}, \phi \circ \mathop{snd}\}} }
  \end{align*}
 \end{definition} 

On the other hand, the public input abstraction $\eta$ deserves some explanation. It can happen that Def.~\ref{sec:aak} fails because the attacker is able to distinguish two input states having the same property $\eta$.
Consider a model $\Model$ of the program $P::= l:= 2l\ast h^2; \Lout{\Sign(l)}$ where $\eta = \Par$ and $\phi = \Idd$. Let $\pi$ be a maximal execution originating from initial state $\sigma$ such that $\sigma(\pi,0)(l)= 2$ and $\sigma(\pi,0)(h)= 1$. Then 
one can find another maximal execution $\pi'$ such that $\sigma(\pi',0)(l)= -2$ and  $\sigma(\pi',0)(h)= 1$. Clearly $\Par(\sigma(\pi, 0)(l)) = \Par(\sigma(\pi',0)(l))$ and $\phi = \True$, while the sign of the outputs are different i.e. $\Sign(4) \neq \Sign(-4)$.  
In \cite{GM04popl} this is called \textit{deceptive} flow, since it only depends on variations of public inputs. However, if one interprets the public input abstraction $\eta$ as secret knowledge that should not be controlled or disclosed to the attacker then
it is reasonable to rule out the program above. Indeed, here the attacker is disclosing  a property stronger than $\Par$ since she observes variations of the sign for inputs of even parity.\\
We now show the equivalence of these definitions and postpone  a further investigation of relation to abstract non-interference as future work.

\begin{proposition}[Equivalence of NANI and AAK]
  For all programs \program{}:
  \begin{displaymath}
    \program{} \models \text{NANI} \; \text{ iff } \; \program{} \models \text{AAK}
  \end{displaymath}
  \begin{proof}
    It is enough to observe that the abstract domain $\rho$ in NANI can be considered as a predicate over public output states. 
    In that case the output action in AAK models the same property. 
  \end{proof}
\end{proposition}
We conclude this section by discussing an interesting example.
\begin{example}
Let $P$ be a program that manipulates a secret variable $h\in{\vec{h}}$, initially known to range over non-negative numbers up to some constant $max$. 
We express this fact by a declassification policy $\phi= 0\leq h \leq max$. Then $P$ is secure since it outputs the same sequence of numbers in every run.
\[P::=  \left [ \begin{array}{ll}
x:=0;  & \\
\while \; (x<h) \; \dow \; \Lout{x}; x++; &\\
\while \; (x<max) \; \dow \; \Lout{x}; x++; &\\
\end{array} \right. \]
Program $P$ satisfies Def~\ref{def:AKD}. Too see this, consider a model $\Model$ of $P$, a maximal execution $\pi$ originating from $\sigma_0 = (max_0, x_0, h_0)$ and any point $i$. $ 0 \leq i \leq \Len(\pi)$. 
Assume $\phi(h_0)$ holds, then for all values $h_i$ such that $\phi(h_i)$, it is possible to find an execution $\pi'$ originating from $(max_0, x_0, h_i)$ and a point $i'$  such that $\TraceOf(\pi, i) = \TraceOf(\pi', i')$. 
In fact, all executions produce a increasing trace of numbers of length at most $max_0$. If $\phi(h_0)$ does not hold then all executions produce the empty trace. 
 
\end{example}

\section{Declassification: Where}

Another well-studied form of declassification regards where in the system sensitive information can be released.
In our framework, the only way to leak secret information is by means of output operations. In particular, any flow of information from a high identifier $h$ to a low identifier $l$ is perfectly fine as long as secret data is not being output. It is irrelevant at which point of a certain epoch the declassification occurs. For this reason, assume that declassification takes place together with the output actions. We model the  release points in the code by special boolean flags $r_e$ initially $false$ and once set to $true$ the program can release the value of expression $e$. Moreover, the flag can no more be updated once it is set to true.
 Assume we are given a set of release points interspersed in the program, say $ \cR_p = \{r_{e_{1}}, \cdots, r_{e_{n}}\}$, and the corresponding release expressions $\cR = \{e_{1}, \cdots, e_{n}\}$ then the goal is to check whether program \program{} leaks more information that what the programmer has already allowed to be disclosed by means of the release points encountered so far. It is worth recalling that our model intends to protect the initial value of secret data, not the current ones. This objective is in line with most other work on noninterference. Let $\powersetOf{\cR}$ be the power set of \cR{} and $\bar{\cE}$ be the complement of \cE{} in \cR{}.
 The formula expressing the absence of attacker knowledge  is given next.
 \begin{definition}[AKR] \label{def:AKR} \mbox{} \\
   Let $\{r_{e_{1}}, \cdots, r_{e_{n}}\}$ be the boolean variables, initially false, serving as flags for the release policy \cR{}.
   A program \program{} satisfies \emph{absence of knowledge modulo release} \cR{} iff:
   \begin{displaymath}
     \program{} \models \globallyOp[]{ \bigvee_{\cE \in \powersetOf{\cR}} \Big( \espmFormula{\vec{l}}{\vec{h}}{\cE} \wedge \bigwedge_{e_i \in \cE}r_{e_{i}} \wedge \bigwedge_{e_j \in \bar{\cE}} \neg r_{e_{j}} \Big) }
   \end{displaymath}
 \end{definition}   
 Note that the conditions above are mutually exclusive with respect to release points, namely given $\pi$ and  $i$, only one formula in the disjunction  holds and that corresponds to the one with release points set to true in execution $\trunc(\pi, i)$.
 \begin{example} \label{ex:release}
   Consider program $P$ with $h_1, h_2\in{\vec{h}}$ and $l\in{\vec{l}}$.
   \begin{center}
     \Lassign{\id[l]{}}{\id[h_1]{}} \!\!\Lseq{}{}
     \Lassign{\id[r_{h_1}]{}}{\val[true]{}} \!\!\Lseq{}{}
     \Lout{\id[l]{}} \!\!\Lseq{}{}
     \Lassign{\id[l]{}}{\id[h_2]{}} \!\!\Lseq{}{}
     \Lassign{\id[r_{h_2}]{}}{\val[true]{}} \!\!\Lseq{}{}
     \Lout{\id[l]{}} \!\!\Lseq{}{}
   \end{center}
   Stores are vectors $(l, h_1, h_2)$ and $\vec{h}$ is the high store $(h_1, h_2)$.
   Intuitively $P$ is secure since the value of a secret is always declassified before being output. Pick $\execution{} \in \model[]{\program{}}$. We show that Def.~\ref{def:AKR} holds for $(\pi,0)$. 
   Initially  $\cE = \emptyset$ is the only candidate such that $\bigwedge_{e_i \in \cE}r_{e_{i}} \wedge \bigwedge_{e_j \in \bar{\cE}} \neg r_{e_{j}}$. It remains to prove that $\execution{}, 0 \models \espmFormula{}{}{\emptyset}$. This trivially holds until the first release point as the trace of any execution up to this point is empty and any execution generates an empty trace at some point.
   Then, we move on to $(\execution{}, 2)$ which is the first execution point after setting the first release flag. At this point, \espmFormula{}{}{\{h_1\}} is required to hold. For the same reason as above, \espmFormula{}{}{\emptyset} holds and by Prop.~\ref{prop:ESPMf-monotonicity} \espmFormula{}{}{\{h_1\}} also holds.
   The trace of $(\execution{}, 3)$ is ``$\mathtt{h}_1$", where $\mathtt{h}_1$ is the initial value of $h_1$, and \espmFormula{}{}{\{h_1\}} is still the formula required to hold. Among all the execution points whose trace is $\mathtt{h}_1$ and whose execution has started with the same initial values for $l$ and $h_1$, there is at least one point whose execution has started with $h_2 = \mathtt{h}_2$ for any $\mathtt{h}_2$. Hence, $(\execution{}, 3)$ satisfies \espmFormula{}{}{\{h_1\}}.
   Similarly, $(\execution{}, 4) \models \espmFormula{}{}{\{h_1\}}$, $(\execution{}, 5) \models \espmFormula{}{}{\{h_1,h_2\}}$ and $(\execution{}, 6) \models \espmFormula{}{}{\{h_1,h_2\}}$. Hence, \program{} satisfies AKR.
 \end{example}

We  now show how Def.~\ref{def:AKR} relates to a similar security condition called \textit{gradual release} \cite{ASgr07}. Although gradual release considers a slightly different computational model, the basic idea is that the attacker knowledge is constant between release points. In the same spirit, we compute the attacker knowledge for a given trace and compare it with the information released over that trace. In particular, if the attacker knowledge is greater than what has been declassified so far, there is an insecure leakage. Given a program $P$, an initial store $\sigma_0$ and a trace $\trace$ originating from that store, we define the knowledge over the trace $\cK(P, \sigma_0, \trace)$ as the set of initial stores that could have led to that trace.
\begin{multline*}
  \cK(P, \sigma_0, \trace) = \\
  \{\sigma\executionPoint[0]{} \mid \exists \executionPoint{} : \sigma\executionPoint[0]{} \approx_{\vec{l}} \sigma_0 \wedge  \TraceOf\executionPoint{} = \executionPointTrace{} \}
\end{multline*}
 As pointed out by Askarov and Sabelfeld \cite{ASgr07}, this set corresponds to the uncertainty of an attacker observing trace $\trace$.

 When reaching a point whose trace is \executionPointTrace{} and execution started in $\sigma_0$, a certain number of release point $r_{\phi}$ have been executed. Let $\cD_{\sigma_0, \trace}$ be the set of common release points that have been executed when reaching any point whose trace is \executionPointTrace{} and execution started in $\sigma_0$ and $\Phi_{\sigma_0, \trace} = \{\phi \mid r_{\phi} \in \cD_{\sigma_0, \trace}\} $. Moreover, let $\cR(P, \sigma_0, \trace)$ be the maximum knowledge authorized, or minimum uncertainty required, at a point whose trace is \executionPointTrace{} for an execution started with the value store $\sigma_0$.
\begin{displaymath}
  \cR(P, \sigma_0, \trace) = \{\sigma \mid \sigma \approx_{\vec{l}} \sigma_0 \wedge \bigwedge_{\phi \in \Phi_{\sigma_0, \trace}} \sigma_0(\phi) = \sigma(\phi) \}
\end{displaymath}
Then, a program is secure if the information disclosed by observing a given trace is less than the information released over that trace; or if the required uncertainty is a subset of the attacker uncertainty.

\begin{definition}[ER] \label{def:ER} \mbox{} \\
  A program \program{} satisfies \emph{epistemic release} iff:
  \begin{displaymath}
    \forall \sigma_0, \trace : \; \cR(P, \sigma_0, \trace) \subseteq \cK(P, \sigma_0, \trace)
  \end{displaymath}
\end{definition}

\begin{example}
  Consider the program in Example~\ref{ex:release} over a boolean domain and $(l, h_1, h_2)$ a triple corresponding to a store. Take $\sigma_0(l) = \True$. Then, for the empty trace $\epsilon$, we have $\cK(P, \sigma_0, \epsilon) = \cR(P, \sigma_0, \epsilon)= \{(\True,\_, \_)\}$.
  Now we pick $\trace = \True$ and $\cK(P, \sigma_0, \True) = \cR(P, \sigma_0, \True)= \{(\True,\True,\_)\}$ since we release $h_1$. Proceeding in this way it is easy to prove that $P$ satisfies ER.
  Suppose that we don't release $h_1$ at the first output. Then we have  $\cR(P, \sigma_0, \True)= \{(\True,\_, \_)\}$ which is clearly not contained in $\cK(P, \sigma_0, \True)$.
\end{example}

\begin{proposition}[Equivalence of AKR and ER]
  For all programs \program{}:
  \begin{displaymath}
    \program{} \models \text{AKR} \; \text{ iff } \; \program{} \models \text{ER}
  \end{displaymath}
  \begin{proof}
    $(\Rightarrow)$ Assume $\program{} \models \text{AKR}$.
    Let $\execution{} \in \model[]{\program{}}$. We show that for all prefixes $\trace$ of $\TraceOf(\execution{}),  \cR(P, \sigma(\pi, 0), \trace) \subseteq \cK(P, \sigma(\pi, 0), \trace)$.
    Consider \executionPoint{} such that $\TraceOf\executionPoint{} = \trace$ and release points $r_{\phi_1}, \cdots, r_{\phi_k}$ being active. By Def.~\ref{def:AKR}, $\pi, i \models \espmFormula{}{}{\cE}$ where $\cE = \{\phi_1, \cdots, \phi_k\}$. Basically, it says that for all \executionPoint[0]{'} such that $\sigma\executionPoint[0]{} \approx_{\vec{l}} \sigma\executionPoint[0]{'}$ and $\bigwedge_{\phi \in \cE} \sigma\executionPoint[0]{}(\phi) = \sigma\executionPoint[0]{'}(\phi)$ (i.e. $\executionPoint[0]{'} \in \cR(P, \sigma_0, \trace)$), there exists \executionPoint{'} such that $\TraceOf\executionPoint{'} = \trace$ (i.e. $\executionPoint[0]{'} \in \cK(P, \sigma_0, \trace)$). This is exactly ER. \\
    $(\Leftarrow)$ Assume $\program{} \models \text{ER}$, we show that $\program{} \models \text{AKR}$.
    Pick any $\execution{} \in \model[]{\program{}}$ and $\executionPoint{} \in \execution{}$. Let $\sigma_0 = \sigma\executionPoint[0]{}$, $\executionPointTrace{} = \traceOf{}\executionPoint{}$ and $\cE = \{\phi_1, \cdots, \phi_k\}$ the set of release whose flag has been set. By Def.~\ref{def:AKR}, AKR requires only \espmFormula{}{}{\cE} to hold at \executionPoint{}. By hypothesis and Def.~\ref{def:ER}, $\cR(\model[]{}, \sigma_0, \executionPointTrace{}) \subseteq \cK(\model[]{}, \sigma_0, \executionPointTrace{})$; therefore, for all \execution{'} such that $\sigma_0 \approx_{\vec{l}} \sigma\executionPoint[0]{'}$ and $\bigwedge_{\phi \in \Phi_{\sigma_0, \trace}}\sigma_0(\phi) = \sigma\executionPoint[0]{'}(\phi)$, there exists \executionPoint{'} such that $\TraceOf\executionPoint{'} = \trace$. As $\cD_{\sigma_0, \trace} \subseteq \cE$, it implies \espmFormula{}{}{\cE}.
  \end{proof}
\end{proposition}

\begin{figure}[htbp]
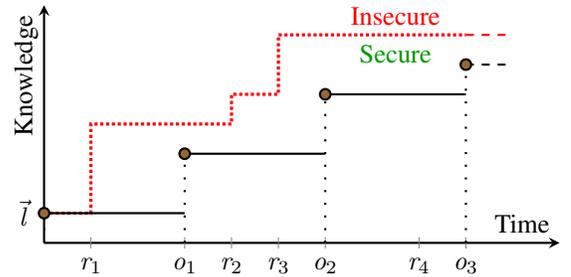

\begin{center}
\ifpdf
  \ifthenelse{\boolean{usePgfplotsFigures}}{
    \begin{tikzpicture}
      \begin{axis}[compat=1.3,
        width=\linewidth, height=30ex, 
        axis x line = bottom, axis y line = left, 
        xmin=0, enlarge x limits={abs=1,upper},
        ymin=0, enlarge y limits={abs=1,upper},
        xlabel=\text{Time}, every axis x label/.style={at={(1,0)},anchor=south east}, 
        ylabel=\text{Knowledge}, every axis y label/.style={at={(ticklabel cs:1)},rotate=90,anchor=north east}, 
        every axis/.append style={thick},
        xtick=\empty, ytick=\empty,
        extra x ticks={1,4,5,8,3,6,9}, extra x tick labels={$r_1$,$r_2$,$r_3$,$r_4$,$o_1$,$o_2$,$o_3$},
        extra y ticks={1}, extra y tick labels={$\vec{l}$},
      ]
    
      \addplot[red, very thick, densely dotted, const plot, no markers] coordinates {(0,1) (1,4) (4,5) (5,7) (8,7) (9,7)};
      \addplot+[red, no markers, dashed] coordinates {(9,7) (10,7)};
    
      \addplot+[black, mark=*, jump mark left] coordinates {
        (0,1) (3,3) (6,5) (9,6)
      };
      \addplot+[black, no markers, dashed] coordinates {(9,6) (10,6)};
      \addplot+[black, loosely dotted, no markers, ycomb] coordinates {
        (0,1) (3,3) (6,5) (9,6)
      };
   
      \node[red,anchor=south] at (axis cs:7.5,7) {Insecure};
      \node[green!60!black,anchor=north] at (axis cs:7.5,7) {Secure};
      \end{axis}
    \end{tikzpicture}
  }{
    \resizebox{\linewidth}{20ex}{\includegraphics{release.pdf}}
  }
\else
 \scalebox{0.6}{ \input{release.pstex_t} } 
\fi
\caption{Knowledge and Release}
\label{figure:example}
\end{center}
\end{figure}

Figure \ref{figure:example} explains  the  epistemic  release wrt. the attacker knowledge. As before, the graphic corresponds to the knowledge about initial secrets that program semantics 
releases by means of the output trace $\trace = o_1o_2o_3$. The black solid line shows how the knowledge can possibly increase in each output point by disclosing information about the secrets. 
The red dotted line shows the secret information declassified in each epoch by release points $r_i$.
Since the dotted line remains above the solid line, the attacker knowledge is less than what the programmer releases by means of these points. Hence the program will satisfy the security condition. 

\begin{example}
Consider a program $P$ (variation of \cite{LZ05}) with $secret, x, y \in {\vec{h}}$ and $in, l \in {\vec{l}}$. $P$ allows a local release point $r_\phi$ with declassification policy
$\phi= hash(h) \modu 2^{64} = in$ i.e. private variable $secret$ can only be leaked comparing the least 64 bits of his hashed value to public input variable $y$.  
 \[P::=  \left [ \begin{array}{ll}
x:=hash(h); y=x\modu2^{64};  & \\
 \ifc \; y=in \; \thenc \; l:=0 \; \elsec \; l:=1;&\\
r_\phi; \Lout{l}; &
\end{array} \right. \]
Applying Def.~\ref{def:AKR}, one can see that for any fixed initial value of identifiers $in, l$, for all initial values $h$ having property $\phi$ the  output value is 1 and all initial $h$ having property $\neg \phi$ 
the output value is  2.
However, if we append to $P$ the following lines of code (where $z \in {\vec{l}}$), it becomes insecure. 
\[P'::= P; z:=x\modu 3; \Lout{z}\]
Indeed, pick $h_1, h_2$ satisfying  $\phi$ and $hash(h_1)\modu 3 \neq hash(h_2)\modu 3$, then it violates the release policy.\\
\end{example}

\section{Declassification: When} \label{sec:declassification-when}
 The last dimension of declassification addressed in this paper is the ``when'' dimension \cite{SSJCS07}. Following an approach similar to the one of Chong and Myers \cite{chong:2004:spfd}, a temporal declassification is a pair $(\phi^C, \phi^D)$ composed of a declassified property $\phi^D$ and a time predicate $\phi^C$ which specifies \emph{when} to declassify $\phi^D$. During any execution, as soon as $\phi^C$ holds, outsiders are allowed to learn $\phi^D$ now and in the future. Let $\Phi$ be a set of \emph{temporal declassifications}, $\Phi^C$ denotes the set of time predicates of $\Phi$ ($\Phi^C = \{\phi^C \mid (\phi^C, \phi^D) \in \Phi\}$) and $\Phi^D$ denotes the set of declassified properties of $\Phi$.
 It has to be noted that there are two types of temporal declassifications. If $\phi^C$ applies to values which are constant during the execution (such as the initial value of a given variable) or are expressed using \initialValueOp{}{} in our model, $(\phi^C, \phi^D)$ describes for which executions an information can be output. A policy stating that a salary can be output only if it is lower than a given constant is an example of such an \emph{inter-execution} temporal declassification. On the other hand, if $\phi^C$ applies to variables whose value vary during the execution then $(\phi^C, \phi^D)$ describes after which event an information can be leaked. An \emph{intra-execution} temporal declassification is for example a policy stating that an information can be provided only after it has been paid for.

 Following the standard definitions of NI (Def.~\ref{def:ONI}) and NID (Def.~\ref{def:NID}), Def.~\ref{def:NITD} formally defines \emph{noninterference modulo temporal declassifications}. It states that at any point \executionPoint{_1} of any execution \execution{_1}, for any execution \execution{_2} started with the same initial public values ($\sigma(\execution{_1}, 0) \approx_{{\vec{l}}} \sigma(\execution{_2}, 0)$) and agreeing on declassifications ($\sigma(\execution{_1}, 0) \approx_{\psi^D} \sigma(\execution{_2}, 0)$) activated so far ($\exists j : \: 0 \leq j \leq i_1 \land \sigma(\execution{_1}, j)(\phi^C)$), there should exists a point \executionPoint{_2} which has the same trace as \executionPoint{_1}.

 \begin{definition}[NITD] \label{def:NITD} \mbox{}\\
   Let $\Phi$ be a a set of \emph{temporal declassifications}, i.e. a set of pairs $(\phi^C_i, \phi^D_i)$.
   A program \program{} satisfies \emph{noninterference modulo temporal declassifications} $\Phi$ iff:
   \begin{multline*}
     \forall \execution{_1}, \execution{_2}  \in \model[]{\program{}} , \forall \executionPoint{_1} \in \execution{_1} : \\
     \left(
       \begin{split}
         & \sigma(\execution{_1}, 0) \approx_{{\vec{l}}} \sigma(\execution{_2}, 0) \quad \land \\
         & \bigwedge_{(\phi^C, \phi^D) \in \Phi}
           \left\{
             \begin{split}
               & \left( \exists j : \: 0 \leq j \leq i_1 \land \sigma(\execution{_1}, j)\phi^C \right) \\
               & \Rightarrow \sigma(\execution{_1}, 0) \approx_{\phi^D} \sigma(\execution{_2}, 0)
             \end{split}
           \right.
       \end{split}
     \right) \\
     \Rightarrow \exists i_2 , \: \traceOf{}\executionPoint{_1} = \traceOf{}\executionPoint{_2}
   \end{multline*}
 \end{definition}

 In our framework, this complex predicate can be naturally expressed using once again the \espmFormula{}{}{} formula. Definition~\ref{def:AKTD} provides the complete epistemic temporal formula that has to hold in order for a program \program{} to satisfy \emph{absence of knowledge modulo temporal declassifications} $\Phi$.

 \begin{definition}[AKTD] \label{def:AKTD} \mbox{} \\
   Let $\Phi$ be a set of \emph{temporal declassifications}.
   A program \program{} satisfies \emph{absence of knowledge modulo temporal declassifications} $\Phi$ iff:
   \begin{align*}
     \program{} \models \; 
     \bigwedge_{\Psi \in \powersetOf{\Phi}} \left( \;
       \weakUntilOp[]{
         \espmFormula{\vec{l}}{\vec{h}}{\Psi^D}
       }{
         \left( \bigvee_{\phi \in (\Phi \setminus \Psi)^C} \phi \right)
       }
       \; \right)
   \end{align*}
 \end{definition}

For any subset of declassification policies $\Psi \subseteq \Phi$, noninterference modulo declassifications $\Psi^D$ (\espmFormula{\vec{l}}{\vec{h}}{\Psi^D}) has to hold until the condition $\phi^C$ of an information not declassified by $\Psi^D$ holds ($\phi^D \notin \Psi^D$). In particular, noninterference (\espmFormula{\vec{l}}{\vec{h}}{\emptyset} by Prop.~\ref{prop:ESPf-eq-ESPMf-empty}) has to hold until the first information is declassified. Generally, if $\Psi^C$ is the set of all declassification conditions which have been triggered so far, noninterference modulo $\Psi^D$ and all superset of $\Psi^D$ has to hold ($\forall \Psi^D_2 \!\!: \espmFormula{\vec{l}}{\vec{h}}{\Psi^D \cup \Psi^D_2}$). However, by Prop.~\ref{prop:ESPMf-monotonicity}, noninterference modulo $\Psi^D$ subsumes noninterference modulo any superset of $\Psi^D$, and is therefore the real policy enforced when the set of conditions triggered so far is $\Psi^C$.

\begin{proposition}[Equivalence of NITD and AKTD] \mbox{} \\
  For all programs \program{}:
  \begin{displaymath}
    \program{} \models \text{NITD} \; \text{ iff } \; \program{} \models \text{AKTD}
  \end{displaymath}
  \begin{proof}
    Let $\Phi_{\executionPoint{}} \subseteq \Phi$ be the set of all temporal declassifications $(\phi^C, \phi^D)$ which have been triggered at execution point \executionPoint{} ($\exists j : \: 0 \leq j \leq i \land \sigma(\execution{}, j)\phi^C$). \\
    \textbf{($\Rightarrow$)}
    For all execution points \executionPoint{_1} and initial stores $\sigma_2^0$ which have the same public values as the initial store of \executionPoint{_1} ($\sigma(\execution{_1}, 0) \approx_{{\vec{l}}} \sigma_2^0$) and agree on $\Phi_{\executionPoint{}}^D$ ($\sigma(\execution{_1}, 0) \approx_{\Phi_{\executionPoint{}}^D} \sigma_2^0$), there exists an execution \execution{_2} started in the initial state $\sigma_2^0$ which has the same trace as \executionPoint{_1} at some point \executionPoint{_2}. This follows from Def.~\ref{def:NITD}, the fact that for all $\phi^C$ not in $\Phi_{\executionPoint{}}^C$ there is no execution point preceding or equal to \executionPoint{_1} such that $\phi^C$ holds, and $\sigma_1 \approx_{\Phi_{\executionPoint{}}^D} \sigma_2$ implies $\sigma_1 \approx_{\phi^D} \sigma_2$ for all $\phi^D$ in $\Phi_{\executionPoint{}}^D$. \\
    The above statement corresponds to: \emph{\espmFormula{\vec{l}}{\vec{h}}{\Phi_{\executionPoint{_1}}^D} holds for all point \executionPoint{_1}} (Def.~\ref{def:ESPM-formula}). All the rest of the proof follows from it. First showing that for any subset $\Psi$ of $\Phi$ and execution point, either \espmFormula{\vec{l}}{\vec{h}}{\Psi} holds (1) or there exists $\phi \in (\Phi \setminus \Psi)$ such that $\phi$ holds in the current execution point or a preceding one (2). Then, AKTD is proved by contradiction. If AKTD does not hold then there exists a subset $\Psi$ of $\Phi$ and an execution point \executionPoint{} such that \espmFormula{\vec{l}}{\vec{h}}{\Psi} does not hold at \executionPoint{}, which would contradict (1), and no $\phi \in (\Phi \setminus \Psi)$ is such that $\phi$ holds in \executionPoint{} or a preceding point, which would contradict (2). \\
    For any $\Psi$, Prop.~\ref{prop:ESPMf-monotonicity} implies that \espmFormula{\vec{l}}{\vec{h}}{\Phi_{\executionPoint{_1}}^D \cup \Psi} holds at \executionPoint{_1}. Hence, for any $\Psi \supseteq \Phi_{\executionPoint{_1}}^D$, (1) holds, and a fortiori (1) or (2). For any $\Psi \in \powersetOf{\Phi}$ not superset of $\Phi_{\executionPoint{_1}}$, there exists $\phi \in \Phi_{\executionPoint{_1}} \setminus \Psi)$ such that $\phi$ belongs to $\Phi \setminus \Psi$ and holds at \executionPoint{_1} or a preceding state. Hence, for any $\Psi \not\supseteq \Phi_{\executionPoint{_1}}^D$, (2) holds, and a fortiori (1) or (2). Therefore, $\text{NITD} \Rightarrow \text{AKTD}$.
    \\
    \textbf{($\Rightarrow$)} The proof follows in the reverse order the same equivalence relations as above; relying on the fact that for any point \executionPoint{_1} $\espmFormula{\vec{l}}{\vec{h}}{\Phi_{\executionPoint{}}^D}$ has to hold.
  \end{proof}
\end{proposition}

\begin{example} Let \program{}, whose code is provided below, be a program that outputs a \id[data]{} after payment of its \id[cost]{}.
  \begin{quote}
    \Lwhile{paid $<$ cost}{\{\Lassign{paid}{paid + note}\}} \!\Lseq{}{} \\
    \Lif{cost $>$ max}{\!\Lout{"\textsf{ok}"}}{\!\Lout{paid}} \!\Lseq{}{} \\
    \Lout{data}
  \end{quote}
 Initial value stores $(\id[paid]{},\id[note]{},\id[max]{},\id[cost]{},\id[data]{})$ are of the shape $(0,\mathtt{n},\mathtt{m},\mathtt{c},\mathtt{d})$ where $\mathtt{n}$, $\mathtt{m}$, $\mathtt{c}$ and $\mathtt{d}$ are integers. The intended security policy is that the \emph{initial} values of $\id[paid]{}$, $\id[note]{}$ and $\id[max]{}$ are public and everything else should be kept secret, except for the \id[cost]{} which can be revealed only if it is not greater than \id[max]{} (note that if \id[cost]{} is not lower than \id[max]{} then the final value of \id[paid]{} must not be revealed either) and \id[data]{} which can be output after payment. In our framework, this policy is formalized by $\id[paid]{},\id[note]{},\id[max]{}\in{\vec{l}}$ and $\Phi = \{(\val[true]{}, \id[cost]{} > \id[max]{}), (\id[cost]{} \leq \id[max]{}, \id[cost]{}), (\id[paid]{} \geq \id[cost]{}, \id[data]{})\}$. The first declassification of $\id[cost]{} > \id[max]{}$ may seem unnecessary, however in order to reveal the cost only if $\id[cost]{} \leq \id[max]{}$ it is required to declassify $\id[cost]{} > \id[max]{}$. Possible traces of \program{} are: ``'' while still paying, ``\textsf{ok}'' and ``\textsf{ok} $\mathtt{d}$'' if $\mathtt{c} > \mathtt{m}$, otherwise ``$x$'' and ``$x$ $\mathtt{d}$'' where $x = \mathtt{n}\times\lceil{}\mathtt{c}\div\mathtt{n}\rceil{}$.
 Obviously, any execution point of \program{} before the first output satisfies noninterference and \espmFormula{\vec{l}}{\vec{h}}{\Psi} for all $\Psi$ (Prop.~\ref{cor:ESPf-subsumes-ESPMf}). However, as the time predicate of $\id[cost]{} > \id[max]{}$ is \val[true]{}, AKTD never requires \espmFormula{\vec{l}}{\vec{h}}{\emptyset} to be satisfied. Only \espmFormula{\vec{l}}{\vec{h}}{\{\id[cost]{} > \id[max]{}\}} is required to be satisfied at the beginning of the execution if $\mathtt{c} > \mathtt{m}$, otherwise \espmFormula{\vec{l}}{\vec{h}}{\{\id[cost]{} > \id[max]{}, \id[cost]{}\}} which is equivalent to \espmFormula{\vec{l}}{\vec{h}}{\{\id[cost]{}\}} as \id[max]{} contains a public data (any executions started with the same public data and \id[cost]{} have to agree on $\id[cost]{} > \id[max]{}$).
 After the loop, payment has been made and $\id[paid]{} \geq \id[cost]{}$ implies that AKTD only requires \espmFormula{\vec{l}}{\vec{h}}{\{\id[cost]{} > \id[max]{},\id[data]{}\}} to be satisfied if $\mathtt{c} > \mathtt{m}$, and otherwise \espmFormula{\vec{l}}{\vec{h}}{\{\id[cost]{} > \id[max]{},\id[cost]{},\id[data]{}\}} which is equivalent to \espmFormula{\vec{l}}{\vec{h}}{\{\id[cost]{},\id[data]{}\}}. If $\mathtt{c} > \mathtt{m}$ then next traces are ``\textsf{ok}'' and ``\textsf{ok} $\mathtt{d}$''. For any initial value store differing only on \id[cost]{} but such that $\id[cost]{} > \id[max]{}$, there exist an execution point whose trace is ``\textsf{ok}'' and another for ``\textsf{ok} $\mathtt{d}$''. For executions where $\mathtt{c} \leq \mathtt{m}$ and after the loop, AKTD only requires that executions started with the same initial value store can generate the same trace. Hence, \program{} satisfies AKTD.
\end{example}

\section{Conclusion and Future Work} \label{sec:conclusion}

We have pointed out a strong connection between temporal epistemic logic and several security conditions studied in the area of language-based security, including (state-based) noninterference and various flavors of declassification. 
We claim that temporal epistemic logic appears to be a well suited logical framework to express and study information flow policies. There have been other attempts at building such general frameworks in the past, including McLean's selective interleaving functions \cite{DBLP:journals/tse/McLean96} and Mantel's modular assembly kit \cite{DBLP:conf/ccs/Mantel05}. These approaches are quite different, and focus more on the modular construction of security properties than their extensional properties. Other notable attempts include Banerjee, Naumann and coauthors work on information flow logics (cf. \cite{DBLP:conf/sp/BanerjeeNR08} involving various specialized constructs to constrain data flow and dependencies between variables. An interesting feature of the epistemic account of information flow is that indirect flows are handled completely indirectly: it is never necessary to explicitly talk about variables on different executions being in agreement, or depending on each other; information flow is fully captured in terms of the effects of these dependencies on agents knowledge.

Our approach is not yet general enough to handle general trace-based conditions. This paper considers programs with output events only, whereas most work on trace-based security conditions address traces consisting of both output and input events. There is no problem in principle to extend our approach to programs with both inputs and outputs, e.g. the interactive programs considered by Bohannon et al \cite{BPSWZ09}. Extending the study in this direction to better understand the role and limits of temporal epistemic definability in security modeling is an important line of inquiry for future work.

The reader will have noticed that we actually use only a very small fragment of the logic we set out to study. For instance, we only use the epistemic possibility operator \possibilityOp[]{} and never its dual \knowledgeOp[]{} (epistemic necessity, knowledge), and never use nesting of epistemic connectives. The former is due to our focus on confidentiality rather than integrity properties.
 Temporal epistemic logic in its standard form may be richer than needed for the application domain; computational or proof-theoretical gains may be made by considering sparser languages.
 Related to this is the general problem of tractability, and if the temporal epistemic setting can be used to develop techniques for more precise information flow analysis.

\bibliographystyle{abbrvnat}

\end{document}